\documentclass[fleqn,10pt]{wlscirep}

\usepackage{graphicx}

\usepackage[utf8]{inputenc}
\usepackage[T1]{fontenc}
\usepackage{amsmath} 
\usepackage{nameref}
\usepackage{hyperref}
\usepackage{booktabs} 
\usepackage{subcaption} 
\usepackage[defaultcolor=red]{changes}

\makeatletter
\def\ttl@useclass#1#2{%
  \@ifstar
    {\ttl@labelfalse\@dblarg{#1{#2}}}
    {\ttl@labeltrue\@dblarg{#1{#2}}}}
\makeatother

\title{
Social aspects of collision avoidance: A detailed analysis of two-person groups and individual pedestrians
}

\author[1,*]{Adrien Gregorj}
\author[1,3]{Zeynep Y\"ucel}
\author[1,2,3]{Francesco Zanlungo}
\author[4]{Claudio Feliciani}
\author[3,5]{Takayuki Kanda}

\affil[1]{Okayama University, Okayama, Japan}
\affil[2]{Osaka International Professional University, Osaka, Japan}
\affil[3]{ATR International, Kyoto, Japan}
\affil[4]{The University of Tokyo, Tokyo, Japan}
\affil[5]{Kyoto University, Kyoto, Japan}

\affil[*]{adrien-gregorj@s.okayama-u.ac.jp}

\begin{abstract}
Pedestrian groups are commonly found in crowds but research on their social aspects is comparatively lacking.  To fill that void  in literature,  we study the dynamics of collision avoidance between pedestrian groups (in particular dyads) and individual pedestrians in an ecological environment, focusing in particular on (i) how such avoidance depends on the group's social relation (e.g.\ colleagues, couples, friends or families) and (ii) its intensity of social interaction (indicated by conversation, gaze exchange, gestures etc). By analyzing relative collision avoidance in the ``center of mass'' frame, we were able to quantify how much groups and individuals avoid each other with respect to the aforementioned properties of the group. A mathematical representation using a potential energy function is proposed to model avoidance and it is shown to provide a fair approximation to the empirical observations. 
    We also studied the probability that the individuals disrupt the group by ``passing through it'' (termed as intrusion). We analyzed the dependence of the parameters of the avoidance model and of the probability of intrusion on groups' social relation and intensity of interaction. We confirmed that the stronger social bonding or interaction intensity is, the more prominent collision avoidance turns out.
    We also confirmed that the probability of intrusion is a decreasing function of interaction intensity and strength of social bonding. Our results suggest that such  variability should be accounted for in models and crowd management in general. Namely, public spaces with  strongly bonded groups (e.g.\ a family-oriented amusement park) may require a different approach compared to  public spaces with  loosely bonded groups (e.g.\ a business-oriented trade fair).
\end{abstract}

\begin{document}

\flushbottom
\maketitle

\thispagestyle{empty}

\section*{Introduction}

\label{sec:Introduction}

Groups represent an important component of pedestrian crowds~\cite{moussaid2010walking,schultz2013group}, and lately they have been the subject of many studies, focusing on such themes as their effect on crowd dynamics~\cite{hu2021social}, the observation of their shape and velocity~\cite{willis2004human,moussaid2010walking,costa2009interpersonal,zanlungo2014potential,zanlungo2015spatial-size,fu2019walking,fu2020dynamic,fu2022analysis,gorrini2016age}, mathematical and computational modeling of interaction dynamics~\cite{zhang2011local,gorrini2016social,zanlungo2014potential,zanlungo2015mesoscopic,karamouzas2014universal,vizzari2013adaptative,koster2011modelling}, and dependence on social structure and interaction level~\cite{yucel2017walk,zanlungo2017intrinsic,yucel2018modeling,yucel2019identification,zanlungo2019intrinsic}. Many works are based on ecological observations~\cite{bandini2014agent,wei2015survey,bandini2014agent, costa2009interpersonal,moussaid2010walking,zanlungo2014potential,zanlungo2015mesoscopic,zanlungo2015spatial-size,zanlungo2017intrinsic,zanlungo2019intrinsic,yucel2018modeling,yucel2019identification,shao2014scene,corbetta2020high,corbetta2018physics}, which may be argued to be indispensable in dealing with social aspects of human behavior~\cite{lui2021modelling,kidokoro2015simulation,fahad2018learning,ono2021prediction,akabane2020pedestrian,kiss2022constrained,costa2009interpersonal}.

To plan safe buildings~\cite{bode2015disentangling,lovreglio2016evacuation, adrian2020crowds,von2017empirical,feliciani2020systematic} and manage crowds in busy public places (i.e.\ transportation hubs) and events~\cite{schultz2010passenger,gayathri2017review,reuter2014modeling}, it is important to consider group behavior in crowd simulators and monitoring/predicting tools~\cite{karamouzas2011simulating,von2016pedestrian,koster2015modelling,he2016proxemic}. This includes taking into account factors such as the structure of different types of groups, their social relations~\cite{feng2016empirical}, and cultural differences~\cite{seitz2017parsimony,chattaraj2009comparison} for more accurate simulations. It is also crucial to model not only the internal dynamics of the group, but also their reaction to the external environment, such as the presence of other pedestrians, while keeping in mind the aforementioned characteristics of each group. Furthermore, it is essential to consider the impact of the group on other pedestrians and how their behavior may be influenced by the presence of the group.

The emergence of autonomous navigating agents in industries like smart vehicles, assistive robots, and drones has led to a significant increase in research attention towards collision avoidance in recent years. Human beings are naturally good at avoiding each other and the famous Shibuya crossing, in Tokyo, is a good illustration of our ability to avoid collision, even in crowded environments. Consequently, researchers tried to model this ability in various ways \cite{zhou2017collision,wang2015new,karamouzas2009predictive}. Early on, in the Social Force Model \cite{helbing1995social}, a repulsive force between particles (representing pedestrians) was used to account for collision avoidance. Such models have yielded good results when used in tandem with path-finding algorithms to navigate autonomous agents \cite{shiomi2014towards}. More recent research has delved into the specific scenario of pairwise avoidance during face-to-face encounters among pedestrians, using real-life trajectory data\cite{corbetta2018physics,corbetta2020high,corbetta2023physics}. They measured the deviation of pedestrians from their expected undisturbed trajectories when encoutering others (1vs1 and 1vsN scenarios) and compared the results to a Langevin-like physics model. They notably showed that interaction with multiple incoming pedestrians is better discribed using non-linear superposition of short-ranged contact avoidance forces.

However, the majority of the studies mentioned above concentrate solely on the collective behavior of groups, i.e.\  the dynamics that drive the group to move as one cohesive unit. Mathematically oriented works on group dynamics may introduce the concept of a  ``group potential'' and examine it by assuming that interactions with pedestrians outside the group can be approximated as white noise~\cite{zanlungo2014potential} or as an external ``mean-field'' potential~\cite{zanlungo2015mesoscopic} on average. Similarly, observation-based works~\cite{zanlungo2015spatial-size,yucel2019identification,zanlungo2017intrinsic,zanlungo2019intrinsic} often describe group properties using probability density functions defined by an average process that neglects the specifics of the environment.

On the other hand, when introducing group behavior into a microscopic simulator, it is essential to incorporate specific rules that describe interactions between the group and the environment, particularly with surrounding pedestrians. Introducing in-group dynamical rules is a logical starting point, and these can simply be added to the collision-avoidance rules used for individuals. For instance, if a two-person group encounters a single-walking pedestrian in an acceleration-based or ``Social Force'' model~\cite{adrian2019glossary,helbing1995social,zanlungo2011social}, the acceleration terms of the group's pedestrians can be obtained by summing the individual-individual collision avoidance term and those resulting from in-group interaction~\cite{zanlungo2020effect}. The behavior of the lone pedestrian can be modeled by adding the two avoidance terms with respect to the pedestrians in the group. In other words, collision avoidance is treated as a one-to-one behavior, group dynamics are regarded as an exclusively in-group phenomenon, and the overall dynamics is the sum of these distinct components.

This is the (classical) linear superposition principle~\cite{Feynman:1494701}, which is typically assumed (and verified) in mechanics and simplifies physical models and their dynamics to a great extent. However,  this principle does not necessarily apply to pedestrian dynamics. For instance, when a single walking pedestrian encounters a two-person group walking side by side, particularly if they are socially interacting, he or she may choose to avoid walking through them, even if it seems like the best choice from a pure collision-avoidance perspective. This intrusion decision would be likely, should the two pedestrians be perceived as independent.

In this work, we investigate a relatively unexplored aspect of pedestrian behavior and crowd dynamics: collision avoidance of (or against) groups, and in particular, its dependence on the groups'   ``social attributes'', which refer specifically to social relation and intensity of interaction.

In doing that, we use two  data sets of pedestrian trajectory including annotations of groups' \textit{social attributes} to investigate the nature of individual-group collision avoidance.  Moreover, we focus particularly on groups composed of 2 people (i.e.\ dyads), since they are much more common than larger ones and they actually constitute their fundamental building block together with triads (i.e.\ for an easy navigation and social interaction, large groups break into sub-groups of 2 or 3 people)~\cite{costa2009interpersonal}. In addition, larger groups (of 3 or more people) may require a categorization of pair-wise social relations or interactions, which may be very complex to formulate or generalize. In that respect, we use the word \textit{group} to refer simply to dyads. As groups' counterpart in collision avoidance, we focus on \textit{individuals}, which is a term we use to refer to people not appertaining to a group.

\section*{Methods}
\label{sec:Methods}

\subsection*{Data sets}
\label{sec:Data sets}

In this study, we used two data sets, namely the ATC data set and DIAMOR data set, both of which are reviewed and approved for studies involving human participants by the ATR ethics board~\cite{zanlungo2015spatial-size,zanlungo2014potential}, are publicly available and contain trajectories derived from range data~\cite{atr2015pedestrian,brscic2013person,glas2014automatic}. From these trajectories, we computed the normalized cumulative density maps of the experiment environments shown in Figure~1.

\begin{figure}[htb]
  \begin{center}
    \includegraphics[width=0.8\textwidth]{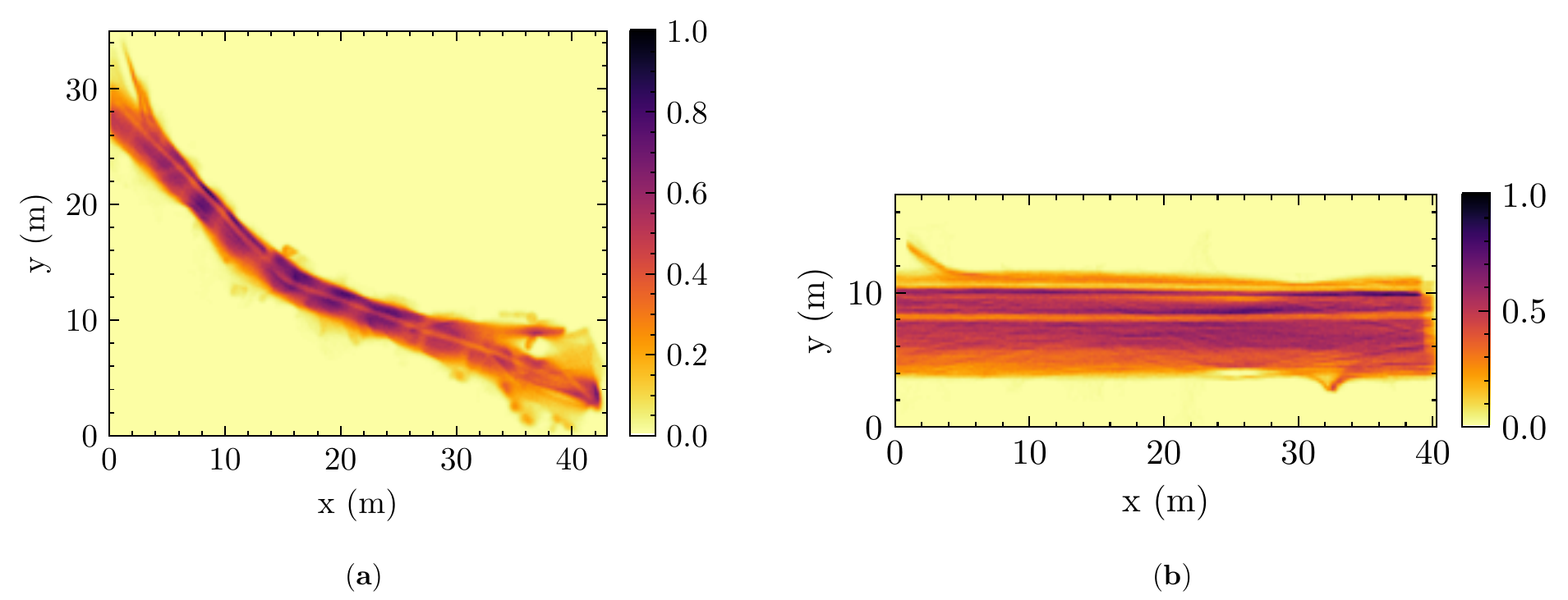}
  \end{center}
  \caption{The normalized cumulative density maps for (a) the ATC data set and (b) DIAMOR data set. The environment is discretized as a 2D mesh with a grid cell size of 10 cm by 10 cm, and the number of observations is counted in each grid cell. Normalization refers to the scaling of this histogram with its maximum value.}
  \label{fig:occupancy_grid}
\end{figure}

The data sets are annotated based on video footage for different social attributes of groups. Specifically, the ATC data set is annotated from the viewpoint of social relations, whereas the DIAMOR data set is annotated from the viewpoint of the intensity of interaction.

For the ATC data set, possible options for social relation are couples, colleagues, family and friends, which are determined  through the domain-based approach of Bugental~\cite{bugental2000acquisition} and correspond to the domains of mating, coalitional, attachment and reciprocal, respectively. This annotation process yields the values presented in Table~1-(a).

For the DIAMOR data set, the intensity of interaction is evaluated at 4 degrees, 0 representing no-interaction and 1, 2, and 3 representing weak, mild and strong interaction, respectively. This annotation process yields the values presented in Table~1-(b). Note that in order not to bias the coders' assessment, we only defined the number of interaction levels as 4, but we did not give any guidelines on what can be considered as weak, mild or strong interaction~\cite{knapp2013nonverbal}. Instead, we let the coders grasp a feeling about different intensities of interaction through a free-viewing task (i.e.\ letting them watch the videos for 3 hours before giving any labels, see Supplementary Information Section 1 for further details.)

\begin{table}[!htb]
  \caption{Number of {groups} annotated with each (a) social relation (in ATC data set) and (b) intensity of interaction (in DIAMOR data set).}
  \label{tab:annotations_stats}
  \begin{center}
    \begin{tabular}{cc}
      (a) & (b)
      \\
      \begin{tabular}{lr}
        \toprule
        Social relation & $\#$ of annotations \\ \midrule
        Couples         & 69                  \\
        Colleagues      & 314                 \\
        Family          & 180                 \\
        Friends         & 253                 \\ \midrule
        Total           & 816                 \\ \bottomrule
      \end{tabular}
          &
      \begin{tabular}{lr}
        \toprule
        Intensity of interaction & $\#$ of annotations \\ \midrule
        0 (no interaction)       & 140                 \\
        1                        & 159                 \\
        2                        & 460                 \\
        3 (strong interaction)   & 100                 \\ \midrule
        Total                    & 859                 \\ \bottomrule
      \end{tabular}
    \end{tabular}

  \end{center}
\end{table}

\subsection*{\color{blue}Approach}
\label{sec:Approach}

Provided that the group and the individual do not perform any collision avoidance, we can expect their (relative) motion to be approximated by a straight line.  We are aware that this assumption requires implicitly the environment to be sufficiently straight and wide (e.g.\ like DIAMOR, see Figure~1-(b) and the discussion in Supplementary Information Section 7) and is valid up to a reasonable range (i.e.\ over a few meters). Namely, in environments with complex geometries (curved or with many obstacles, intersections etc.), the pedestrians need to deviate as part of their interaction with the boundaries. Similarly, over long distances, they will eventually meet some walls, or divert towards different goals, making their relative motion bent. Nevertheless, in a sufficiently straight corridor  and on a scale of few meters, we can expect it to be a good approximation.

This trivial assumption can be considered to serve as a hypothesis, opposite to what we actually anticipate. Based on such a hypothesis, the deviation of (relative) motion from a straight line can be attributed to group-individual collision avoidance. Specifically, by measuring this deviation with respect to different social relations or intensities of interaction (of the group), we may understand the reflections of such group attributes on collision avoidance.

\begin{figure}[htb]
  \begin{center}
    \includegraphics[width=0.8\textwidth]{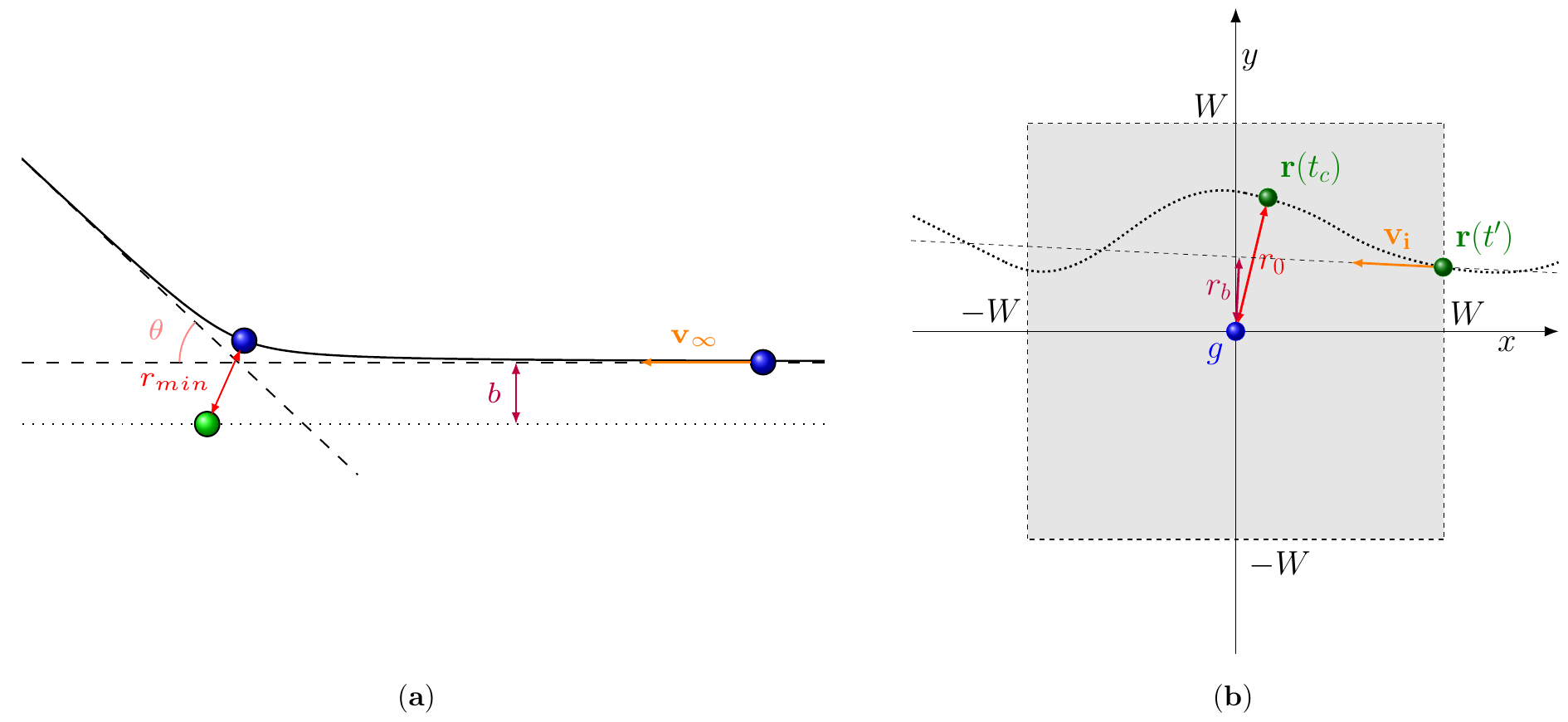}
  \end{center}
  \caption{(a) Illustration of the scattering problem in physics. A mobile particle (in blue) is projected toward a fixed particle (in green). The impact parameter, $b$, is the straight-line distance between the particles, and $r_{min}$ is the closest approach. The particle is deflected with an angle $\theta$. (b) A typical pedestrian avoidance situation in the \textit{group-centered reference frame}. The individual enters the vicinity of the group (gray region) at time $t^{\prime}$ with velocity $\mathbf{v}_i$. The straight-line distance from the individual to the group is denoted by $r_b$. At time $t_c$, the individual is closest to the group at a distance of $r_0$.}
  \label{fig:scattering}
\end{figure}

This formulation presents a striking resemblance to one of the fundamental problems of Physics, namely, the ``scattering problem'', where a ``particle'' (blue ball in Figure~2-(a)) is shot on a ``target'' (green ball), and its deviation from the straight line motion is used to study the interaction potential. In the original scattering problem (see Figure~2), this deviation is assessed by accounting for the straight-line distance $b$ (called the impact parameter) and closest approach $r_{\text{min}}$, which is derived from the  scattering angle $\theta$, as an accurate measurement of  particles' location is very difficult. By repeating the experiment with different impact parameters and estimating the corresponding closest approach $r_{\text{min}}$, one can get an approximation for the potential acting on the particle.

In this study, we establish a simple duality relation between the above-mentioned problem and our group-individual collision avoidance scenario. Namely, the impact parameter $b$ is replaced with a \textit{straight-line distance} $r_b$ and the closest approach can simply be measured as the shortest distance $r_0$ between pairs of trajectory data points of the group and the individual. In Section~\nameref{sec:Observables} we elaborate in detail on how we define these observable quantities.

Using this approach from physical sciences to describe human behavior represents obviously a strong approximation, not only because human behavior is too complex to be modeled through simple physical forces, but also because it completely ignores the effect of the environment, which, in physical parlance, is equivalent to a strong and non-uniform external force. Nevertheless, as we will see, this approach still allows us to grasp the fundamentals of the collision dynamics between groups and individuals and to  quantify the interaction. At this point, it is also worth stressing that the proposed model in Section~\nameref{sec:Modeling} is aimed at assessing the effect of different social attributes in a qualitative way, rather than reproducing quantitatively human behavior.

\subsection*{Observables}
\label{sec:Observables}


\subsubsection*{Data preparation and transformation}
\label{sec:Data preparation and transformation}

We first carry out a data preparation step by (i) removing atypical/non-characterizing motion (waiting, running etc.), (ii) representing the group as a single unit (its geometrical center) and (iii) focusing on frontal encounters of groups and individuals, for which we expect the pedestrians to be able to judge the social attributes of the incoming party.

Subsequently, we transform trajectories of the group  $g$ and the individual  $i$ to a reference frame, which is co-moving with the group. Namely, at each time instant (i) the positions of the group and the individual $\mathbf{r}_{g,i}$ are translated such that the group (center of mass) is positioned at the origin and (ii) their velocities $\mathbf{v}_{g,i}$ are rotated such that the velocity of the group is directed towards $x^+$. Finally, the velocities $\mathbf{v}_{g,i}$ are translated by $-\mathbf{v}_{g}$, rendering the group immobile. The main purpose of this transformation is to provide an easier visualization of relative position in 2D, which represents the position of the individual with respect to the group center, while having the group motion as a preferential direction. On the other hand, most of our analysis is based on the absolute value of the relative distance between the group center and the individual, which is rotationally invariant and independent of frame choice.


\subsubsection*{Relative distance {r}}
\label{sec:Relative distance {r}}

Similar to Corbetta et al.\cite{corbetta2018physics}, our analysis is based on $\mathbf{r}$, the relative position between the group center and the individual,
\begin{equation}
  \mathbf{r}(t)=\mathbf{r}_{i}(t)-\mathbf{r}_{g}(t).
\end{equation}
Its time derivative is the relative velocity $\mathbf{v}$,
\begin{equation}
  \mathbf{v}(t)=\mathbf{v}_{i}(t)-\mathbf{v}_{g}(t).
\end{equation}
The absolute value (norm) of $\mathbf{r}$ is simply denoted as $r$.


\subsubsection*{Straight-line distance $r_b$}
\label{sec:Straight-line distance $r_b$}

The straight-line distance $r_b$ is computed as the shortest distance from the origin (i.e.\ translated position of the group) to the line, which passes through the point at which the individual enters a pre-defined vicinity around the group termed as \textit{window of observation}.  This refers to the area in the group-centered reference frame from $-W$ to $W$ meters both along $x$ and $y$ axes (i.e.\ along the group's motion direction and the direction orthogonal to that). Empirically a $W$  of 4~m is seen  to contain the most significant part of the group-individual collision avoidance (see Figure~2-(b) for an illustration. Refer to previous literature~\cite{cinelli2008locomotor, kitazawa2010pedestrian} and Supplementary Information Section 3 for details on the choice of $W$).

Let $t^{\prime}$ be the time instant at which the individual enters $W$ and let $\mathbf{r}(t^{\prime})$ be its relative position at that instant. According to the hypothesis mentioned in Section~\nameref{sec:Methods}, provided that there is no collision avoidance the individual will follow a path starting at $\mathbf{r}(t^{\prime})$ and move along its velocity vector at that instant $\mathbf{v}(t^{\prime})$. In this case, the straight-line distance $r_b$ can be computed as the shortest distance between this line and the origin (i.e.\ translated position of the group),
\begin{equation}
  r_b
  =
  \frac
  {
    ||\mathbf{r}(t^{\prime})
    \times
    \mathbf{v}(t^{\prime})||
  }
  {
    || \mathbf{v}(t^{\prime})||
  }.
  \label{eq:rb}
\end{equation}
In the analysis, in order to alleviate the impact of orientation noise on the velocity of the individual, we averaged its velocity vector over 4 time instants (before $t^{\prime}$) and used this mean velocity in Equation~\ref{eq:rb} instead of $\mathbf{v}(t^{\prime})$.

\subsubsection*{Observed minimum distance $r_0$}
\label{sec:Observed minimum distance $r_0$}

The observed (i.e.\ actual) minimum distance $r_0$ between the group and the individual is simply,
\begin{equation}
  r_0
  =
  \min_t
  \big(
  r(t)
  \big)
  = r(t_c),
  \label{eq:r0}
\end{equation}
where  $t_c$ is the time instant at which the individual is closest to the group. The time steps, at which pedestrian positions are recorded, are obviously discrete. Nevertheless, in order to have a more accurate estimation of $r_0$, one can also interpolate $\mathbf{r}(t)$ between two consecutive time steps $t_k$ and $t_{k+1}$ by using the velocity vector at time $t_k$,
\begin{equation}
  \mathbf{r}(t)\approx \mathbf{r}(t_k)+(t-t_k)\mathbf{v}(t_k),\;t\in[t_k,t_{k+1}).
\end{equation}
This procedure allows detecting minimum distances not only \textit{exactly at sampling instants}, but also \textit{at intermediate time points} between consecutive samples, which yields a much more accurate estimation of $r_0$ (refer to Supplementary Information Section 4 for details).

\subsubsection*{Scaled distances}
\label{sec:Scaled distances}

Groups' interpersonal distance is shown to depend on their social relation and interaction intensity~\cite{yucel2019identification,yucel2018modeling}. Thus, we represent the distances defined above in two ways: in a group-independent way (in meters) and in a group-dependent way, in which the unit of distance is the average interpersonal distance of dyads with the given social bonding~\cite{zanlungo2017intrinsic}). In the text, we denote distances measured in meters with the normal font (e.g.\ $r$) and scaled distances measured in interpersonal distance units with a bar (e.g.\ $\bar{r}$). Since we observed that results concerning scaled values are in general easier to interpret, in the main text we mostly report those (for further details, refer to Supplementary Information Section 5).

\subsection*{Analysis of collision avoidance}
\label{sec:Analysis}

As mentioned in Section~\nameref{sec:Methods}, our study of the collision avoidance dynamics between groups and individuals is fundamentally based on examining the relation between $r_b$ and $r_0$. The relation between these two observables, although defined in a slightly different way, has been studied in a similar way for 1-1 encounters by Corbetta~\cite{corbetta2018physics}. In what follows we define two different methods to analyze this relation and then propose a method to model it.

\subsubsection*{Empirical relation between $r_b$ and $r_0$ and its statistical analysis}
\label{sec:Empirical relation between bar rb and bar r0 and its statistical analysis}

To examine the distribution of $r_b$ versus $r_0$, the values of $\bar{r}_b$ are quantized into bins of 0.5~unit and for each bin, the average and standard error of the corresponding values of $\bar{r}_0$ are computed. The choice of  0.5 as bin size was primarily driven by empirical observations. For certain combinations of distance $r_b$ (or $\bar{r}_b$) and bonding (social relation or interaction level), setting a smaller bin size results in having bins with little or no data. Conversely, using a larger bin size  decreases the resolution. In that respect, we consider a bin size of 0.5 to strike a balance between these competing factors. The results will be presented and discussed in Section~\nameref{sec:Results on the relation between bar r0 and bar rb}.

\subsubsection*{Intrusions}
\label{sec:Intrusions}

Small groups, such as dyads, have been shown to usually prefer deviating to avoid splitting~\cite{zhang2022walking}. Nonetheless, we found situations where the individual passes through the group (i.e.\ between group members), and we refer to them as ``intrusion''. For simplicity's sake, we define the probability of intrusion as the probability of having $r_0$ smaller than the group interpersonal distance (see Section~\nameref{sec:Scaled distances}). We perform a statistical analysis to investigate the dependence of intrusion on the social attributes of the group. The results will be shown and discussed in Section~\nameref{sec:results_intrusion}, whereas the details of the computational procedure can be found in Supplementary Information Section 6.

\subsubsection*{Modeling}
\label{sec:Modeling}

Many models of pedestrian collision avoidance are based on ``Social Forces''~\cite{adrian2019glossary,helbing1995social,zanlungo2011social}, which may be defined through a potential. It has
been reported that using position-dependent potentials in modeling of pedestrian collision avoidance fails to reproduce detailed behavior. Even if we assume that a ``Social Force'' approach may reproduce actual human behavior, the corresponding potential should at least be velocity-dependent and based not on current distance but on future distance at the moment of predicted closest approach~\cite{zanlungo2011social,karamouzas2014universal}. Nevertheless,  determining a potential that may, at least qualitatively, describe the collision avoidance between groups and individuals, represent an important first step towards a more realistic quantitative modeling.

Let us first review how we can study the potential energy between two \textit{interacting} bodies in physics (note that while discussing the physical model, we use the word ``interaction'' to refer to the effect that the bodies exert on each other.).
The study of such a ``scattering'' problem is obviously a cornerstone of physics, and the non-Quantum formalism analyzed in this section was used to study such important problems as the structure of atoms~\cite{landau2000mechanics} and  gravitational lens effect due to space-time curvature~\cite{blau2011lecture} among others.
In general, the ``bodies'' in focus are very complex and composed of many particles (e.g.\ planets, stars). Nevertheless, due to the scale of the problem, they may be treated as point particles themselves (in our ``pedestrian scenario'', the group will be represented with a single point).

Their interaction is determined by a potential energy $U(\mathbf{r})$, which is in general a function of only relative position (a result connected to invariance under space translations and equivalent to Newton's third law~\cite{hall2013quantum}). Nevertheless, in many important applications, the potential is central, i.e.\ rotationally invariant, and depends only on the magnitude of the distance, $U(r)$.

In such a case, it is shown that the interesting (potential-dependent) dynamics is studied in the $r$ variable~\cite{landau2000mechanics}. Defining the reduced mass $\mu$ as
\begin{equation}
  \mu
  =
  \frac{m_1 m_2}{m_1+m_2},
  \label{eq:reduced_mass}
\end{equation}
where $m_1$ and $m_2$ are the masses of the two bodies,
the angular momentum $\mathbf{L}$ and energy $E$ result to be constants of motion,
\begin{equation}
  E
  =
  \frac{\mu}{2}\dot{r}^2+\frac{||\mathbf{L}||^2}{2\mu r^2}+U(r).
  \label{eq:energy}
\end{equation}

In a scattering problem, the system is not bound and $r$ diverges for $t\to \pm \infty$. In most physical applications, we can only measure the scattering angle and the velocity far before/after the interaction. We can then compare the measured angle with a theoretical result involving an integral. However, if the full trajectory is known, a simpler way to study the system is available.

We study the system far before interaction, i.e.\ for $t\to -\infty$ and $r \to +\infty$, and call the corresponding asymptotic speed $v_{\infty}$. We see that the absolute value of angular momentum can be written as
\begin{equation}
  L\equiv ||\mathbf{L}||=\mu v_{\infty} b,
  \label{eq:angular_momentum}
\end{equation}
where the impact parameter $b$ is the minimum value of $r$ assumed in case of straight motion (i.e.\ no interaction). Assuming $\lim\limits_{r\to +\infty} U(r)=0$, we obtain
\begin{equation}
  E_{\infty}=\frac{\mu}{2}v_{\infty}^2.
\end{equation}
On the other hand, since we actually have interaction, the minimum distance $r_{\text{min}}$ in the observed trajectory turns out to be different than $b$, i.e.\ $r_{\text{min}}\neq b$. At $r=r_{\text{min}}$, having a minimum, we have $\dot{r}=0$ and the corresponding energy is
\begin{equation}
  E_0=\frac{\mu v_{\infty}^2 b^2}{2 r_{\text{min}}^2}+U(r_{\text{min}}).
\end{equation}
Conservation of energy implies $E_{\infty}=E_0$ and provides the following relation for the value of $U(r)$ at $r=r_{\text{min}}$
\begin{equation}
  \label{eq:potential}
  U(r=r_{\text{min}})=\frac{\mu v_{\infty}^2}{2}\frac{r_{\text{min}}^2-b^2}{r_{\text{min}}^2}.
\end{equation}
This relation enables studying the potential $U(r)$, provided that $v_{\infty}$, $b$ and $r_{\text{min}}$ are measured. In modeling collision avoidance between pedestrians based on the above framework, we assume that
\begin{equation}
  \label{eq:repulsive}
  \frac{\mathrm{d}U(r)}{\mathrm{d}r}<0\; \forall r \Rightarrow U(r)>0 \; \forall r.
\end{equation}
In other words, the force is assumed to be repulsive. Namely, denoting $\mathbf{F}_1$ as the force acting on body 1, and recalling
the usual definition $\mathbf{r}=\mathbf{r}_1-\mathbf{r}_2$, we have
\begin{equation}
  \label{eq:repulsivef}
  \mathbf{F}_1=-\boldsymbol{\nabla}U(r)=-\frac{\mathrm{d}U(r)}{\mathrm{d}r}\frac{\mathbf{r}}{r}.
\end{equation}

We apply these physical concepts in a pedestrian scenario to model the ``collision avoidance potential'' between groups and individuals. As mentioned in Section~\nameref{sec:Methods}, $r_b$ is inspired by the impact parameter $b$, whereas $r_0$ corresponds to the closest approach $r_{\text{min}}$. Thus, the term $v_{\infty}$ in Equation~\ref{eq:potential} should be approximated by using the relative velocity when $r_b$ is computed. But since pedestrian velocities have a small variation, we may consider it to be almost constant. In a similar way, as usual when studying
``forces'' that determine the pedestrians' cognitive decisions, all masses are considered to be equal (to one)~\cite{adrian2019glossary} and we may remove $\mu$ from the equation.
Finally, since the approach is completely of a qualitative nature, we opt for ignoring the overall constant in Equation~\ref{eq:potential} and study the following simplified version,
\begin{equation}
  \label{eq:potential2}
  U^\prime(r=r_0)
  =
  \frac{r_0^2-r_b^2}{r_0^2},
\end{equation}
to which we will refer to as the ``collision avoidance potential'' (defined as a dimensionless pure number).

A comment on Equation~\ref{eq:potential2} is probably needed. This equation does not represent the functional form of the dependence of the potential on $r$. Instead, it shows which is the value of $U$ attained at $r_0$ {\it given that the straight-line distance is $r_b$}. Different values of $r_b$ allow us to probe different values of $U$, where the smaller $b$ is, the higher $U'$ is. Nevertheless, Equation~\ref{eq:potential2} clearly allows us only to probe values $U'<1$. This is due to the fact that in the computation of $U'$
the value of the initial kinetic energy
\begin{equation}
  \frac{\mu v_{\infty}^2}{2}
\end{equation}
is taken as fixed, and we are measuring the probed values of the collision avoidance potential as multiples of such kinetic energy. Note that in particle physics short distances are indeed probed by using very high kinetic energies.

The results are shown and discussed in Section~\nameref{sec:results_potential}, whereas the details of the computational procedure are described in Supplementary Information Section 7. In addition, in Section~\nameref{sec:Comparison to individual-individual collision avoidance} we also show the results concerning a similarly defined potential describing individual-individual collision avoidance.

\section*{Results and discussion}
\label{sec:results}

\subsection*{Results on relative frame pdfs}
\label{sec:Results on relative frame pdfs}

The \textit{group-centered reference frame} is particularly suitable to study the 2D distribution of $\mathbf{r}$, i.e.\ of the position of the individual around the group.
Figures~3 and 4 show the 2D distributions in relation to different social relations and interaction intensities of the group, respectively,
using as a distance unit the groups' average interpersonal distance. Note that, in order to highlight the specificities of each social attribute \textit{as compared to the whole}, we depict the difference between a given attribute and the overall 2D average, which  is computed as an unweighted average of the distributions of all relating cases. Therefore, positive values depict an increased likeliness of presence for the individual, while, reciprocally, negative values depict a decreased likeliness.

\begin{figure}[htb]
  \begin{center}
    \includegraphics[width=0.8\textwidth]{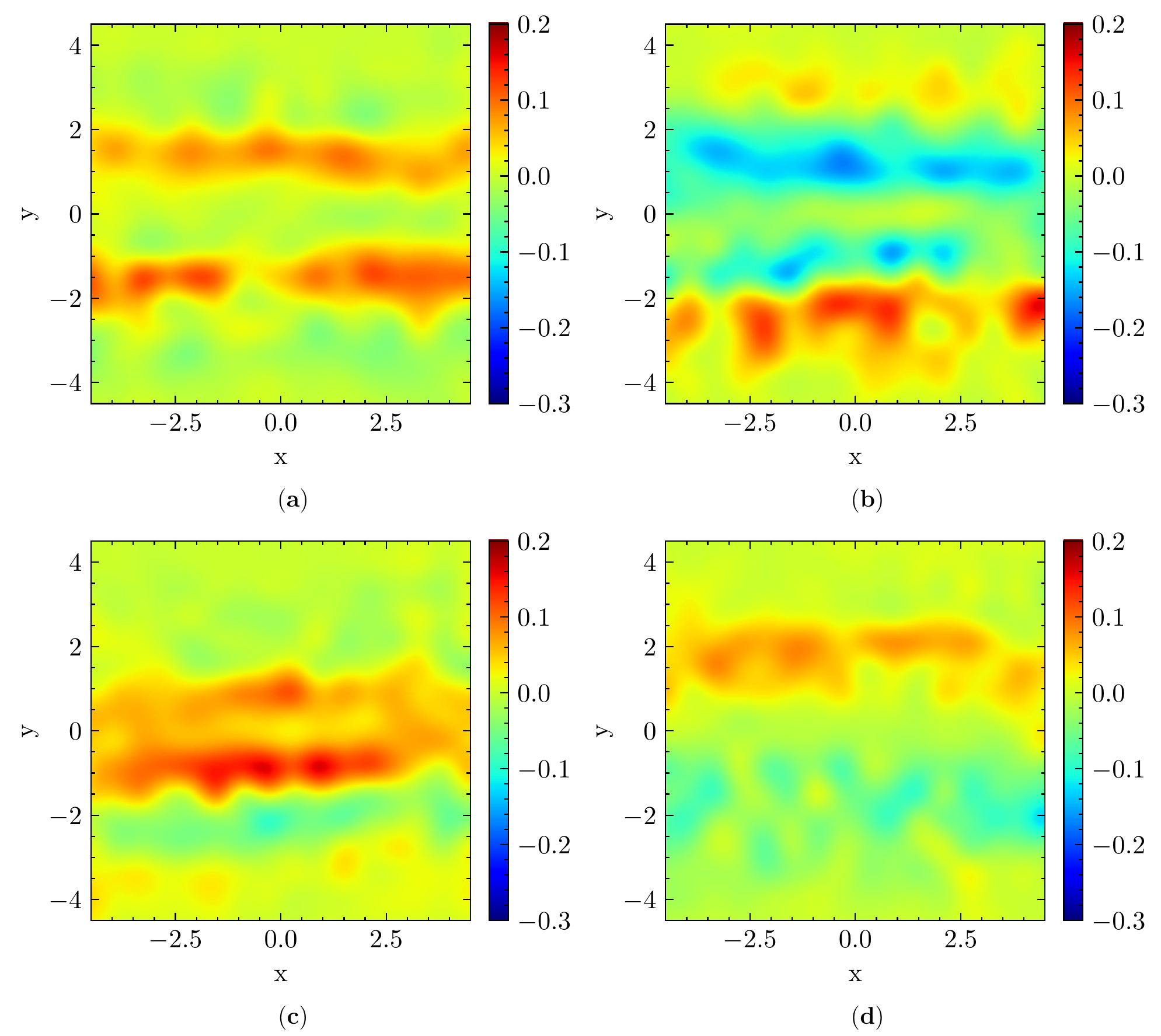}

  \end{center}
  \caption{2D probability distribution of individuals' position $\bar{\mathbf{r}}$ relative to overall average. Positions are shown in the group-centered reference frame and the $x$ axis is aligned with the direction of motion of the group. Each sub-figure depicts the difference between the distribution relating to a certain social relation and an unweighted average concerning  all social relations. (a) Colleagues, (b) couples, (c) families, (d) friends. The color scales  are adjusted for highlighting the differences.}
  \label{fig:2D_pdf_atc}
\end{figure}

The effects of varying social relations are presented in Figure~3. Comparing Figure~3-(a) with Figure~3-(b) and (d), one may notice that individuals do not have a prominent preference to pass on the right or left side of colleagues, whereas they prefer to pass more on the right for couples and on the left for friends (as compared to the overall average). In addition, they pass with a very small distance ($r \approx 0$) more often for families (see Figure~3-(c)) than for other kinds of social relations, which may be due to a more dispersed configuration of family group members~\cite{yucel2019identification}. On the other hand, in Figure~3-(b) we see very clearly two low probability horizontal stripes, roughly located around $y=\pm 1$. As these stripes  correspond more or less   to  group members' positions, they suggest that the
group's abreast formation is rarely disturbed in couples.

\begin{figure}[htb]
  \begin{center}
    \includegraphics[width=0.8\textwidth]{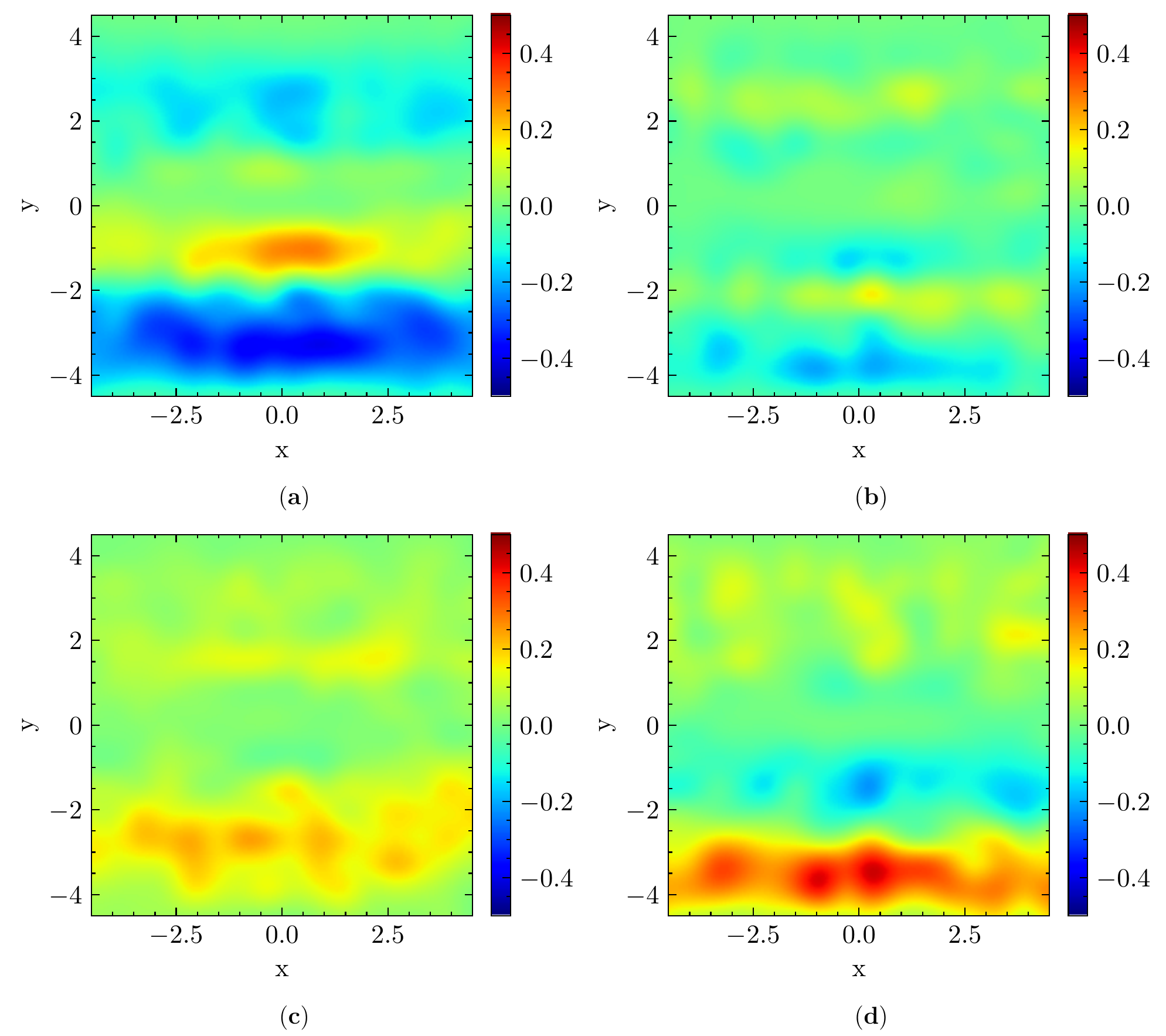}
  \end{center}
  \caption{2D distribution of individuals' position $\bar{\mathbf{r}}$ relative to overall average. Positions are shown in the group-centered reference frame and the $x$ axis is aligned with the direction of motion of the group. Each sub-figure depicts the difference between the distribution relating to a certain intensity of interaction and an unweighted average of all intensities. (a) 0, (b) 1, (c) 2, (d) 3.  The color scales  are adjusted for highlighting the differences.}
  \label{fig:2D_pdf_diamor}
\end{figure}

Concerning social interaction, the difference with respect to varying intensities is much more noticeable, the most interesting one being between 0 and 3 (see Figures~4-(a) and (d)). Namely, concerning groups annotated as non-interacting (i.e.\ with 0 intensity of interaction), the center stripe presents positive values, while the lower and upper stripes $y \approx \pm 2.5$  present negative values, indicating that individuals are more likely to maintain a trajectory directly facing the group (possibly even intruding it) (see Figure~4-(a)). Reciprocally, from Figure~4-(d) we can see that individuals are less likely to position themselves on a colliding trajectory with the group and prefer to place themselves on its side, when it has a high intensity of interaction. There are  interesting  left/right asymmetries in Figure~4, which  may be related to the tendency of Japanese pedestrians to move mainly on the left, and overtake on the right~\cite{zanlungo2012microscopic}. This tendency may cause low-interaction groups, when
they are not intruded on, to have a relatively higher possibility to be passed on their left than on their right, since
they are expected to have a higher speed than highly interacting ones. We do not have a clear interpretation for the right/left asymmetry between the couples distribution
in Figure~3-(b) and the friends distribution in Figure~3-(d).

\subsection*{Results on the relation between $\bar{r}_0$ and $\bar{r}_b$}
\label{sec:Results on the relation between bar r0 and bar rb}

\begin{figure}[htb]
  \begin{center}
    \includegraphics[width=0.8\textwidth]{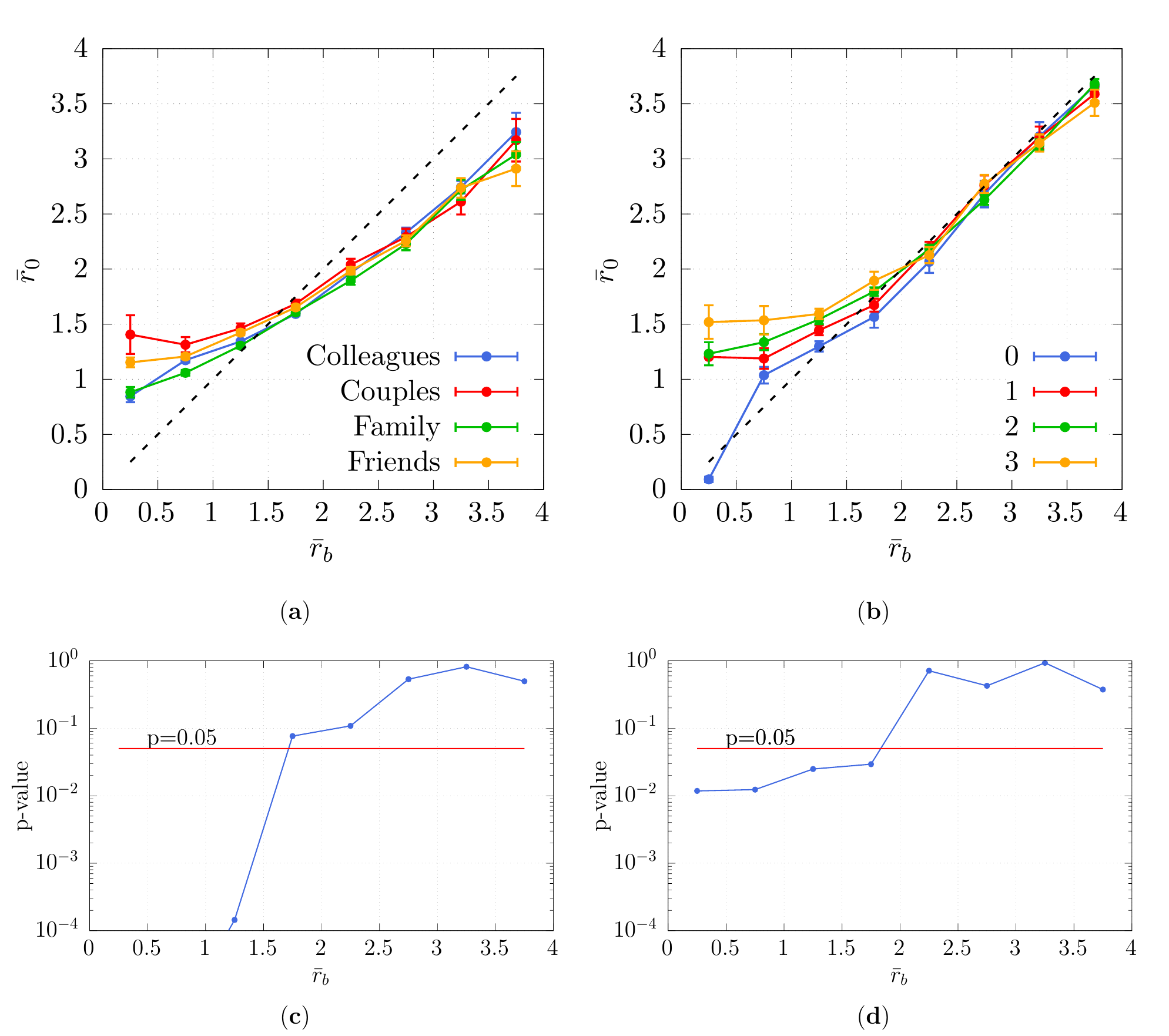}
  \end{center}
  \caption{Observed minimum distance $\bar{r}_0$ as a function of the undisturbed straight-line distance $\bar{r}_b$ (a) for various social relations and (b) intensities of interaction of the group. Error bars report standard error intervals. The dashed line corresponds to the $\bar{r}_0=\bar{r}_b$ linear dependence.
    $p$-values for the ANOVA of $\bar{r}_0$ (c) for various social relations and (d) intensities of interaction of the dyad. In (c), results for {$\bar{r}_b<1$} are not displayed as very low values were obtained ($p < 10^{-6}$).}
  \label{fig:rb_r0_rb_r0_pvalues}
\end{figure}

We divide the range of $\bar{r}_b$ into bins of 0.5~unit and compute the mean and {standard error} of $\bar{r}_0$ corresponding to  each bin. The results are depicted in Figures~5-(a) and (b) and are similar to previous results\cite{corbetta2018physics}.
Smaller values of $\bar{r}_b$ indicate that the straight line trajectory of the individuals would require them to pass very close to the group. In addition, $\bar{r}_b < 0.5$ signifies a distance smaller than half of the group interpersonal distance, which means that the individual would need to \textit{intrude on the group} (if moving straight).

Concerning social relations, we observe that when $\bar{r}_b < 0.5$, the average value of  $\bar{r}_0$ is considerably larger for couples and friends than for colleagues and families (see Figure~5-(a)). In other words, there is a strong resistance against intruding on groups with the former social relations. Concerning intensity of interaction, we have a similar observation for higher intensities of interaction (from 1 to 3, see Figure~5-(b)), while for 0 intensity  we have $\bar{r}_0\approx \bar{r}_b$ for
all $\bar{r}_b$ values.

These observations make us believe that the social attributes of the group do impact group-individual collision avoidance. Specifically, there is larger avoidance, when there is a strongly-bonded group involved (i.e.\ couples, friends or with high intensity of interaction).

The statistical significance of these results can be assessed through an ANOVA (see Supplementary Information Section 8 for considerations regarding the necessary assumptions). To that end, we compute the $p$ values concerning each bin shown in Figures~5-(a) and (b) and demonstrate the results in Figures~ 5-(c) and (d), respectively. Regarding lower values of $\bar{r}_b$ (i.e.\ $\bar{r}_b < 1.5$), we observe statistical significance (i.e.\ $ p < 0.05$) concerning both social relation and intensity of interaction. Regarding larger values of $\bar{r}_b$ (i.e.\ $\bar{r}_b > 2$), there is no statistically significant difference, as it can be expected observing the overlapping curves in the corresponding regions of Figures~5-(a) and (b).

\begin{figure}[htb]
  \begin{center}
    \includegraphics[width=0.8\textwidth]{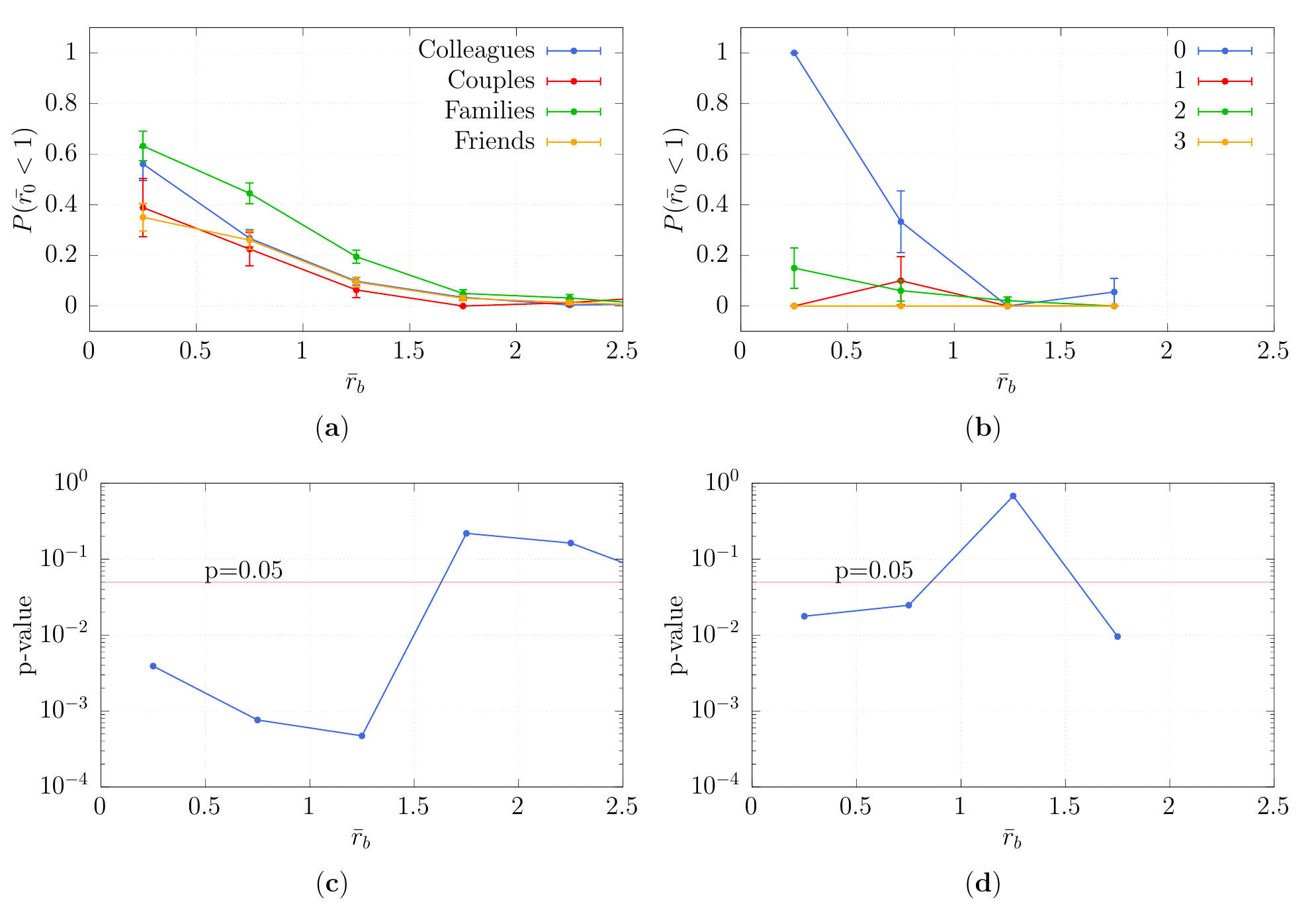}
  \end{center}
  \caption{Probability that the distance $\bar{r}_0$ is smaller than 1 for (a) for various social relations and (b) intensities of interaction of the dyad. Pearson's $\chi^2$ p-values for the hypothesis of independence of the frequencies of samples verifying $\bar{r}_0 < 1$  for (c) for various social relations and (d) intensities of interaction (of the group).}
  \label{fig:probabilities_probabilities_pvalues}
\end{figure}

Let us also notice that in Figure~5-(b) concerning the DIAMOR data set, for $\bar{r}_b \gg 1$, we have $\bar{r}_b \approx \bar{r}_0$ regardless of intensity of interaction, in agreement with the hypothesis that collision avoidance can be ignored for such values (Equation~\ref{eq:potential2}). The fact that this is not the case in ATC, where we actually observe $\bar{r}_b > \bar{r}_0$ for $\bar{r}_b \gg 1$, is considered to be an effect of the ATC environment being less straight and narrower (see Figure1-(a)).

To compensate for this effect in the computation of the potential, we perform a linear correction in the computation of $\bar{r}_b$ in Section~\nameref{sec:results_potential}. The details of this correction are presented in Supplementary Information Section 7. In addition, results concerning the relation between $r_b$ and $r_0$, i.e.\ values measured in meters and not scaled with group interpersonal distance, are shown in Section~\nameref{sec:Comparison to individual-individual collision avoidance}.

\subsection*{Intrusion}
\label{sec:results_intrusion}

It is noticeable that the observed minimum distance $\bar{r}_0$ reaches particularly low values in some encounters. For instance, the first bin in Figure~5-(b) for intensity of interaction 0 presents an average value of $\bar{r}_0$ smaller than 1. This means that the distance from the center of mass of the group to the individual gets smaller than the group interpersonal distance (see Supplementary Information Section 5). In such cases, it is likely that the individual is actually intruding on the group instead of deviating, essentially following the straight line trajectory.

To quantify the frequency of such intrusions, we computed the probability of $\bar{r}_0$ being smaller than 1. Specifically, this is an empirical probability computed as the ratio of the number of observations with $\bar{r}_0 < 1$ to the total number of observations (for a given bin of $\bar{r}_b$). The results are shown in Figures~6-(a) and (b). Here, we see that there is indeed a correlation between the probability of intrusion and the social bonding of the group being intruded on. Namely, individuals have a higher probability to intrude on loosely-bonded groups (i.e.\ colleagues, families and non- or slightly-interacting groups) than strongly-bonded groups (couples, friends and strongly interacting groups).

The statistical significance of this observation is assessed through  Pearson's $\chi^2$ test and the relating $p$-values are presented in Figures~6-(c) and (d). The difference in probability of intrusion concerning different social relations is significant ($p < 0.05$), when $\bar{r}_b$ is smaller than 1.5. On the other hand, for the intensity of interaction we have a significant difference of intrusion for $\bar{r}_b < 1$.

Actually, the average distance of a group member from the group center is $\bar{r}_0=0.5$. The corresponding analysis for the probability of having $\bar{r}_0<0.5$ is shown in Supplementary Information Section 6.

\subsection*{Potential}
\label{sec:results_potential}

\begin{figure}[htb]
  \begin{center}
    \includegraphics[width=0.8\textwidth]{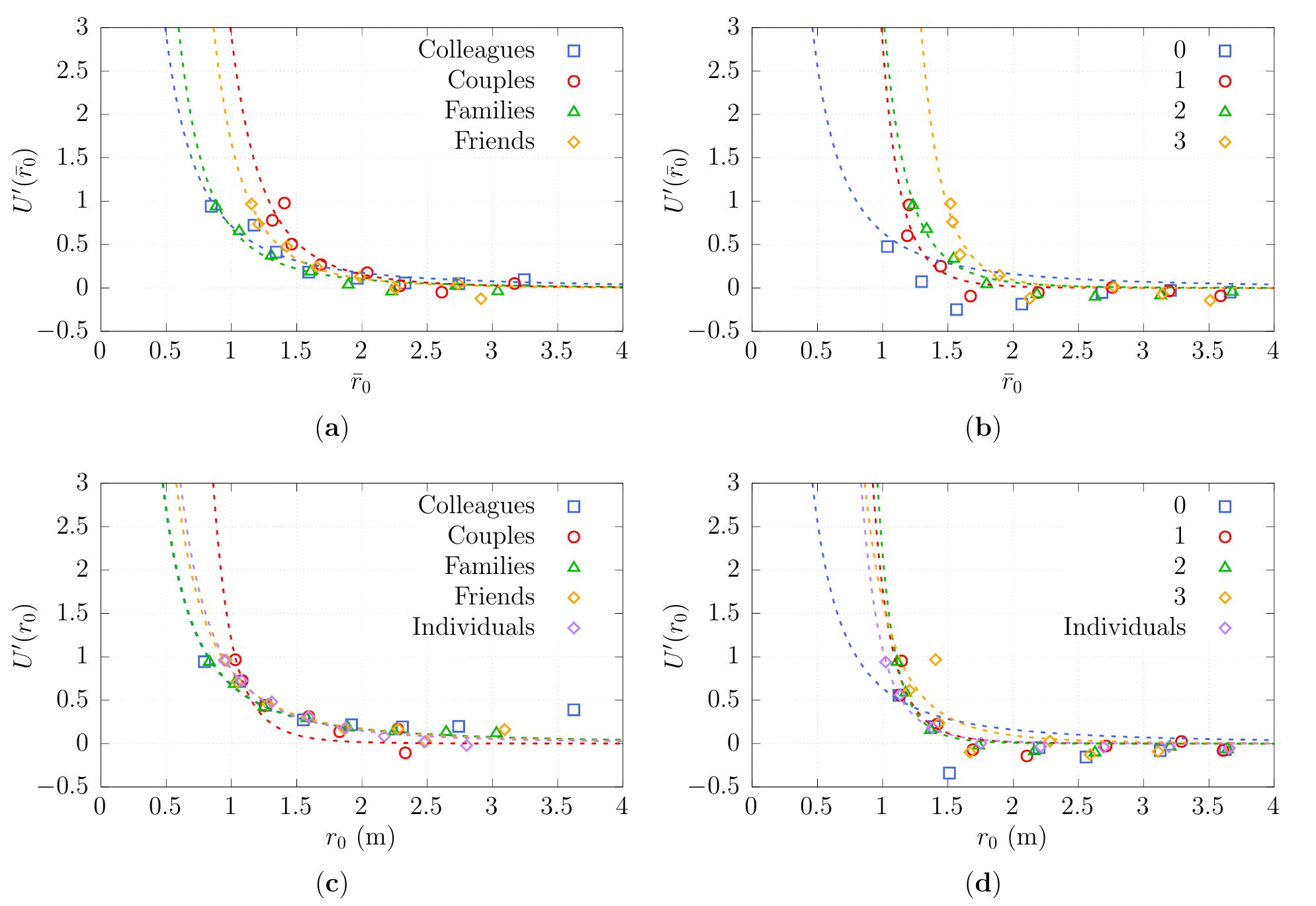}
  \end{center}
  \caption{Collision avoidance potential $U^\prime (\overline{r}_0)$ (a) for various social relations and (b) intensities of interaction of the group. Dashed lines correspond to a power function fit of the quantized data.
    Collision avoidance potential  $U^\prime (r_0) $ (c) for various social relations and (d) intensities of interaction of the group. Dashed lines correspond to an exponential fit of the quantized data. (c) and (d) report a comparison to individual-individual (non-group) interaction using non-scaled distances.}
  \label{fig:potential_potential_ng}
\end{figure}

As described in Section~\nameref{sec:Modeling}, we study $U^\prime (r_0) $ (see Equation~\ref{eq:potential2})
to model the ``potential'' representing the interaction between the group and the individual. To that end, we again quantize the values of $\bar{r}_b$ and compute the corresponding mean values of $\bar{r}_0$ before calculating the values of the potential $U'(\bar{r}_0)$ for each bin. Figures~7-(a) and (b) show the relating values. Additionally, to extrapolate outside the range available, a function of the form ${k}/{r^\beta}$ is fitted to the data using non-linear least squares, illustrated with dashed lines in Figures~7-(a) and (b).

Interestingly, the potential is shown to be affected  by the nature of  the social bonding of the group. As a matter of fact, stronger bondings (e.g.\ couples, high intensity of interaction) generate a ``stronger potential'' (i.e.\ with a steeper negative derivative) which, as seen in Sections~\nameref{sec:Results on the relation between bar r0 and bar rb} and \nameref{sec:results_intrusion} ``causes''  individuals to deviate more, and significantly decreases their probability to intrude on the group.
On the other hand, loosely-bonded groups (e.g.\ colleagues and non- or slightly-interacting groups) generate a weaker potential, resulting in a smaller deviation and a higher chance of intrusion.

The discussion above concerns results obtained using distances scaled with the group interpersonal distance; results concerning computations performed using distances measured in meters are shown in Section~\nameref{sec:Comparison to individual-individual collision avoidance}.

\subsection*{Comparison to individual-individual collision avoidance}
\label{sec:Comparison to individual-individual collision avoidance}

Many practitioners simulate crowds on the basis of individuals. Thus, it is interesting to compare the above-mentioned potentials with results obtained for individual-individual interaction. The results (using distances measured in meters, i.e.\ not scaled
by group interpersonal distance) are shown in Figures~7-(c) and (d).

Note that groups are larger (than individuals)  and expected to exert a stronger ``social force'', but they are also susceptible to being disrupted and intruded on  (passed at $\approx 0$ distance to their geometrical center). Also, while it is expected (statistically) that collision avoidance between individuals is symmetric, it may be that groups interact less than individuals by deviating very little. These effects seem to balance and potentials for collision avoidance between individuals are quite similar to those with groups.

Nevertheless, it may be seen that potentials describing low intensity social interactions, colleagues and families have typically a less steep derivative than the one for individual-individual encounters, while the opposite is observed for high-intensity social interactions and (in particular) for couples.

\section*{Conclusion}
\label{sec:conclusion}

In this work, we analyzed how group-individual pedestrian collision avoidance depends on the group's social relation and social interaction intensity.

In detail, we verified that when straight motion (i.e.\ absence of collision avoidance) would lead to a possible collision, the actual  minimum distance $r_0$ between the individual and the group  is a growing function of social interaction intensity, and assumes a higher value for couples and friends. Similarly, individuals have a stronger tendency to ``intrude'' or ``disrupt'' a group by passing at a distance comparable to the group interpersonal distance when they face groups with low interaction intensity and colleagues and families, as can be verified both by studying 2D distance probability distributions, and by performing a statistical analysis on the probability that the minimum distance becomes smaller than the group interpersonal distance.

We also introduced a ``potential'' to study the dependence of ``intensity of collision avoidance'' on relative distance, by mimicking the theoretical modeling of two-body scattering in classical mechanics. This approach, which may be used as a guiding light in the development of a ``social force model'' of individual-group interaction, shows again that the potential determining collision avoidance tends to grow much faster with decreasing distance values (i.e.\ it has a steeper negative derivative) for strongly interacting groups, couples and families.

The latter result is particularly clear when studied using the group's average interpersonal distance as a length unit. A further comment on this result may be necessary, since the tendency of individuals not to pass through ``strongly bonded dyads'' (such as couples, friends and strongly interacting dyads) may be due not only to some kind of ``social rule'', but also to the fact that passing through these groups is actually harder due to the narrower space between them.

To this respect, we should finally comment also on the results concerning families, which may be a little counter-intuitive by suggesting that families are somehow perceived as weakly interacting and are often ``intruded''~\cite{pouw2020monitoring}. It should be stressed that, as reported by Zanlungo et al.~\cite{zanlungo2017intrinsic}, the families in the ATC data set are mostly composed of parent-child pairs, that often do not walk abreast, or at least have a weaker tendency to walk abreast. The authors of the original study justify  this tendency by referring to ``the erratic behavior of children'', but it may also be related to a stronger hierarchical structure in a parent-child dyad with respect to couples, friends and colleagues~\cite{zanlungo2017intrinsic}. It may thus be argued that the tendency of individuals to approach families at a shorter distance may depend on families being less spatially structured, or correspondingly having a higher tendency to change their spatial structure. Such role of group spatial structure in individual-group interaction could be the subject of future studies, possibly when larger data sets collected in more suitable environments will be available.

We believe  that our results and inferences point out interesting  variabilities in pedestrian motion due to social aspects of human navigation~\cite{rojas2016dynamic}. A valuable implication of our study is that infrastructure design could be adapted to the nature of the social bonding of its users. We can speculate that, for instance, if a particular environment is known to be frequented mostly by strongly bonded groups, such as an amusement park, providing additional space (e.g.\  by widening corridors or walkways) to allow for collision avoidance may make it more comfortable. Nevertheless, these qualitative considerations should ultimately be corroborated with quantitative simulation models that include our findings.  By taking into account the social dynamics of the people using a particular space, designers and architects could create environments that are more conducive to safe and efficient movement. This could help to reduce the risk of accidents and improve the overall user experience. We also hope that  using models which account for the expected social composition of the crowd may help in improving the performance of tracking and simulation systems~\cite{templeton2015mindless}.

\section*{Data availability}

The data sets analyzed during the current study are available at \url{https://dil.atr.jp/ISL/sets/groups/}.

\bibliography{citations}

\section*{Acknowledgements}

This work was supported by the JSPS Grant-in-Aid for JSPS Fellows 22J20686.
This work was financially supported by the JST-Mirai Program Grant Number JPMJMI20D1 and the JSPS KAKENHI Grant Number JP20K14992.
This work was (in part) supported by JST Moonshot R and D under Grant Number JPMJMS2011, Japan.

\section*{Author contributions statement}

A.~G.\  took part in  Data curation, Formal analysis, Investigation, Methodology, Software,  Validation, Writing – original draft, Writing – review \& editing.
Z.~Y.\ took part in  Conceptualization,  Formal analysis, Data curation, Formal analysis, Methodology,  Supervision,  Writing – original draft, Writing – review \& editing.
F.~Z.\ took part in  Conceptualization,  Formal analysis,  Methodology, Writing – original draft, Writing – review \& editing.
C.~F.\ took part in   Funding acquisition, Project administration, Writing – review \& editing.
T.~K.\ took part in  Funding acquisition, Project administration, Writing – review \& editing.
All authors reviewed the manuscript.


\section*{Additional Information}
The authors declare no competing interests.

\appendix

\section{Specifics of experiments and data}
\label{sec:Specifics of experiments and data}

In this section, we give details about experimental conditions and specifics of data and annotations.

\subsection*{ATC data set}
\label{app:ATC data set}

The ATC data set is a large pedestrian data set, which is introduced by Zanlungo et al.~\cite{zanlungo2015spatial-size} and has since been used in several studies, especially those focusing on social aspects of crowd dynamics~\cite{lui2021modelling,kidokoro2015simulation,fahad2018learning,ono2021prediction,akabane2020pedestrian,kiss2022constrained} and it is freely available~\cite{atr2015pedestrian}. It has several properties, which make it particularly fit for the investigation of such factors and we will point them out as we explain the specifics of the data set below.

The recording location is the ground floor of a multi-purpose center, which involves offices, shops and facilities for special events such as expositions. In particular,
the tracking area consisted of an atrium and a corridor which, along with being bordered by shops, serve also as a passage between a train station, a ferry terminal and the
office area. Therefore, in addition to accommodating the visitors of the multi-purpose center, it is also constantly frequented by commuters. In that respect, the recorded pedestrians come from a diverse social background (e.g.\ age group, occupation etc).

The recording area is quite large, i.e.\ approximately 900 m\textsuperscript{2}. Therefore, it allows continuous tracking of individuals for long distances (i.e.\ up to 50~m), which helps also in the annotation of social relations. Note that in the analysis we discarded the trajectory segments collected in the atrium and used the segments collected in the corridor illustrated in Figure~\ref{fig:occupancy_grid}-(a), which is over 40~m long.

The recording time span is more than 800 hours within a one year time window. In such, it involves workdays, weekends, national holidays, public festivals, and special days of the business center such as exhibitions and trade shows. These occasions lead to an increase in density and a diversity in the purpose of visit.

The raw data are in specific depth and video information registered by an RGBD sensor network. Based on the depth information and the algorithm of Br{\v{s}}{\v{c}}i{\'c} et al.~\cite{brscic2013person}, the pedestrians are tracked automatically (on the 2D floor plane) and the resulting trajectories can freely be downloaded~\cite{atr2015pedestrian}. The concerning normalized cumulative density map of the environment is found as in Figure~\ref{fig:occupancy_grid}-(a). Note that the cumulative density map is basically a 2D histogram. We consider the environment as a 2D mesh with a grid cell size of 10 cm by 10 cm and count the number of observations in each grid cell. The normalization refers to the scaling of this histogram with its maximum value.

Note that since the sampling frequency of the sensors is quite high (i.e.\ 20 Hz), tracked trajectories are influenced by gait dynamics. Furthermore, if velocities were simply computed as discrete differences between the tracked trajectory points, the relatively small error in position due to sensor noise would be magnified. For this reason, Zanlungo et al.\ proposed   re-sampling the data by averaging pedestrian positions over 0.5 s time windows, in order to reduce the influence of pedestrian gait and other sorts of measurement noise, as well as to assure uniform sampling in time, and also in this work we are going to use such re-sampled data~\cite{zanlungo2015spatial-size}.

Based on the video information, the pedestrians are annotated according to group membership and several intrinsic group features~\cite{zanlungo2017intrinsic}. Specifically, human coders labeled which pedestrians constitute a social group and which pedestrians move alone.
In addition, for the groups they also annotated intrinsic features such as age, purpose of visit etc. One specific feature, which we focus on in this study, refers to the apparent social relation of the group members. The possible options are couples, colleagues, family and friends, which correspond to the domains of mating, coalitional, attachment and reciprocal defined by Bugental~\cite{bugental2000acquisition}, respectively.
(See Table~\ref{tab:annotations_stats}-(a) for the outcome of the annotation process).

\subsection*{DIAMOR data set}
\label{app:DIAMOR data set}

The DIAMOR data set was introduced by Zanlungo et al.~\cite{zanlungo2014potential} and used particularly for the purposes of group recognition and motion modeling~\cite{brscic2017modelling, glas2014automatic,zanlungo2014potential}.

The recording location is an underground pedestrian street network in a commercial district of Osaka, Japan. The entire underground network is composed of a total of more than several
kilometers of walking path, and it is connected to the Osaka-Umeda railway and underground station complex, which is considered to be one of the busiest in the world (the busiest outside Tokyo) and visited daily by millions of pedestrians. For example, according to the Osaka municipal transportation bureau, the three metro stations located in the underground area had a daily number of passengers of  more than 700K in the fiscal year 2019. The DIAMOR data set includes recordings from a junction of two straight corridors in a relatively peripheral portion of this street network and we focus on one of these. Similar to ATC data set, several train stations, business centers, shopping malls etc.\ are accessible from the recording location leading to diversity in pedestrian profile.

The recording area is roughly 200 m$^2$ and allows continuous tracking along approximately 50~m.
The recording time span is two weekdays and a total of eight hours of recordings are available, which, although shorter than the ATC data set, we consider to be enough for the purposes of this study.

Similar to the ATC data set, it is composed of depth and video information. The depth information is used to derive the trajectories of the pedestrians based on the method reported by Glas et al.~\cite{glas2009laser} and the trajectories can be freely downloaded~\cite{atr2015pedestrian}. As a result of this tracking process, the normalized cumulative density map shown in Figure~\ref{fig:occupancy_grid}-(b) is obtained.

Based on the video information, human coders were asked to annotate groups and individuals (people who do not belong to a group). Of course, it is not wrong to say that each group member is an ``individual'', but within the context of this study we refer to them explicitly as ``group member'', and use the word ``individual'' specifically to refer to people who do not belong a group. Coders also annotated  whether or not members of dyads were engaged in interaction (oral communication, possibly accompanied with non-verbal elements such as gestures or gaze exchange, as defined by Knapp et al.~\cite{knapp2013nonverbal}), and the corresponding intensity of interaction (evaluated at 4 degrees from 0 representing no-interaction to 3 representing strong-interaction). Note that the annotations of ATC data set are inherently disjoint, i.e.\ the annotation labels can be considered to be mutually exclusive nominal variables. On the other hand, the annotation labels of the DIAMOR data set can be viewed as ordinal variables (i.e.\ with a gradual relation). The outcome of the annotation process (i.e.\ the number of observations for each intensity of interaction) is summarized in Table~\ref{tab:annotations_stats}-(b).


\section{Data preparation}
\label{sec:Data preparation}

Both the ATC  data set and the DIAMOR data set are collected in an ecological environment. Namely, they contain trajectories collected from uninstructed people moving freely. Although  they were not recorded secretly (i.e.\ there were signboards informing the pedestrians that a data collection campaign was being carried out),  the pedestrians' awareness of being recorded is anticipated to have a negligible effect on how they move, in particular as compared to participant experiments performed in artificial (laboratory) environment.
In that respect, since the data are collected under uncontrolled settings, the tracked trajectories may contain behaviors like waiting, running etc.
From the point of view of this study, such cases are not of interest. In order to eliminate atypical/non-characterizing observations, each trajectory is treated as explained below.

Let a group (dyad) be described as an unordered pair composed of (two members) $p$ and $q$, i.e.\ $g = (p,q)$ and let $i$ denote an individual.
For the sake of simplicity, we reduce a group to a single mobile agent in the data preparation phase. Namely, the location of the group is represented by the \textit{group center of mass} $\mathbf{r}_g$ and its velocity is represented with \textit{group velocity} $\mathbf{v}_g$. Specifically, concerning a group $g = (p,q)$, $\mathbf{r}_g$ and $\mathbf{v}_g$ are represented as the average positions and velocities of $p$ and $q$ at each time step, respectively.

\begin{equation}
  \begin{split}
    & \mathbf{r}_g = \frac{\mathbf{r}_p + \mathbf{r}_q}{2}, \\
    & \mathbf{v}_g = \frac{\mathbf{v}_p + \mathbf{v}_q}{2}.\\
  \end{split}
  \label{eq:grp_cm_vel}
\end{equation}

Thereby, we treat a group $g$ and an individual $i$ in the same manner and first check the sufficiency of the number of trajectory data points. Provided that $\mid \left\{ \mathbf{r}_{g,i}\right\}\mid \geq 16$, the trajectories are considered to have enough data points for characterizing locomotion. Note that this corresponds to a minimum of 8 seconds of observation, since the sampling time step is 0.5~s. Any trajectory which includes fewer samples is discarded.

Next, we check instantaneous speeds $ {{v}}_{g,i}$ and remove the trajectories which are associated with too low or too high speeds (i.e.\ out of walking range). For judging the typical speed range of pedestrians, we referred to the literature on human locomotion. Based on the results reported by Zanlungo et al.~ \cite{zanlungo2014potential}, $g$ and $i$ are considered to depict typical walking motion, if their instantaneous speed lies within the range $0.5 \leq v_{g,i} \leq 3$ (in m/sec). Otherwise, they are assumed to be ``not walking'' and discarded.


As mentioned in
Section Introduction of the main track, we focus on the effect of social attributes of (the group) on collision avoidance. To have an understanding of those, the peers need to have sufficient visual information about each other. Therefore, we start with conditioning on having \textit{a frontal view of each other}, which implies moving in opposite directions. In other words, if $g$ and $i$ move in the same direction, one party will be leading and the other following, such that the leading party will not see and thus not be aware of the other, and the following party will have limited information about the leading one (e.g.\ on social relation, age, interactions, etc.). In theory, the group and individual might also approach each other at an angle, but in the studied environments such cases are rare and we do not consider them in this work. However, if they move in opposite directions, they will have the opportunity to watch the incoming party and get a sense of its social features.

For ensuring a frontal view, we detected the relative motion direction of $g$ and $i$ and considered only those $g$ and $i$, which move in opposite directions.
Let $\phi$ represent the angle between the velocity vectors $\mathbf{v}_g$ and $\mathbf{v}_i$ at a given time instant,
\begin{equation}
  \phi
  =
  \arccos(\mathbf{v}_g \cdot \mathbf{v}_i) / (||\mathbf{v}_g||\,||\mathbf{v}_i||).
  \label{eq:phi}
\end{equation}
Then, $g$ and $i$ are considered to be moving in opposite directions, if $3\pi/4 \leq \phi < \pi$.
Note that, in addition to boasting a bigger potential from the viewpoint of our purposes, in the studied ``bi-directional'' environments, considering  opposite relative motion direction   has also the advantage that the number of observations associated with it is significantly higher than those with other directions (namely, $68\%$ of all observations), which is preferable for the statistical analysis performed in our study.


\section{Window of observation}
\label{sec:Window of observation}

\begin{figure}[htb]
  \begin{center}
    \includegraphics[width=0.75\textwidth]{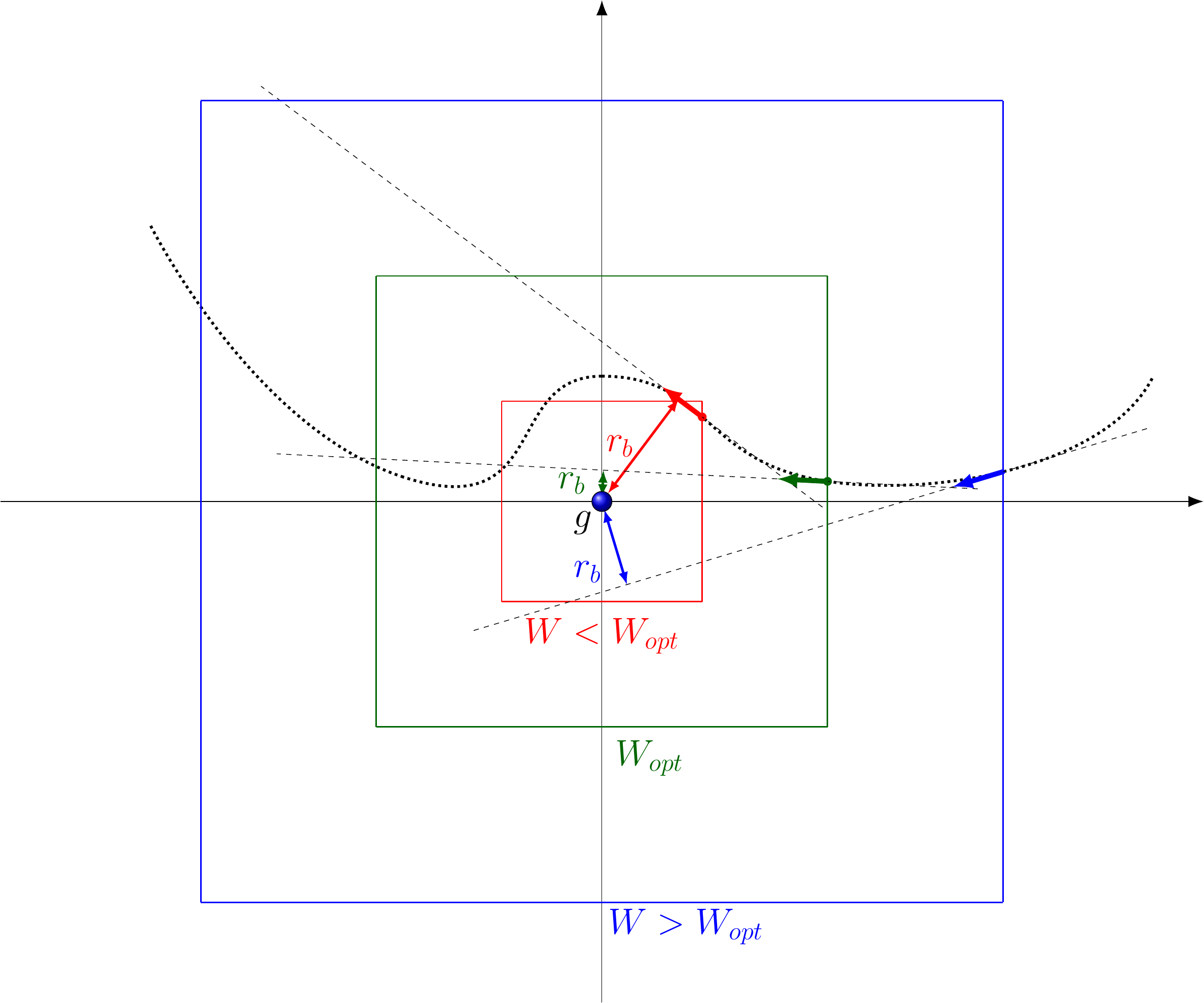}
  \end{center}
  \caption{Illustration of the impact of the choice of the window size $W$ on the computation of the straight-line distance $r_b$. In red, a small window violating \textit{completeness}, in blue, a large window violating \textit{atomicity} and in green, an "optimal" window which satisfies both constraints. Note that
    since we study   frontal encounters in the group reference frame, individuals will always enter the window from the right.
  }
  \label{fig:windows}
\end{figure}

As measures at infinity are obviously not feasible, in order to measure the
straight-line distance $r_b$ we need to define a window of observation (centered on the group). In particular, this window should satisfy two competing properties, \textit{completeness} and \textit{atomicity}. On the one hand, the window needs to be large enough to verify that, when entering and exiting from the side of the window\, the individual is not yet (significantly) influenced by the group (i.e.\ he/she is not yet engaged in an avoidance maneuver). This guarantees that the deviation is entirely contained in the window of observation, hence the notion of \textit{completeness}. On the other hand, the window should not be too large, so that the assumption that the individual would walk on a straight line in the absence of the group stays valid. Indeed, although pedestrians certainly do not walk on straight lines at all times, at a relatively small scale and in the absence of perturbations (resulting from either external sources such as the environment or other pedestrians, or internal sources such as a change of planned destination) it can be expected that one's trajectory will be close to a straight line (\textit{atomicity}). Figure~\ref{fig:windows} illustrates the impact of the size of the window.

Regarding \textit{completeness}, we referred to literature on collision avoidance and searched for a reasonable threshold value.  Cinelli and Patla found that, the ``safety zone'', i.e.\ the area in which individuals allow a moving object to approach before initiating an avoidance behavior, is on average 3.73~m~\cite{cinelli2008locomotor}. Furthermore, Kitazawa et al.\ showed that pedestrians gaze most at other approaching individuals, when they are on average 3.97~m away, and that they seldom look at pedestrians at longer distances than this~\cite{kitazawa2010pedestrian}. Therefore, we deliberated that 4~m is a reasonable lower bound for considering that their mutual influence is still null.

Regarding \textit{atomicity}, the environment needs to be taken into consideration. In a straight and wide corridor, like in DIAMOR, the straight line assumption is more  founded than for a more complex environment, like the bent and narrow corridor in ATC where pedestrian have to follow the curve and will have naturally less straight trajectory. Nonetheless,  the discussion provided in
Section~\ref{sec:Linear correction to rb}
suggests that  the  disparity due to environment geometry can be accounted for with a linear correction. Therefore, we argue that the value of 4~m derived from the \textit{completeness} condition is adequate also to satisfy the atomicity condition.


\section{Improving accuracy of estimation for the observed minimum distance $r_0$}
\label{sec:Improving accuracy of estimation for the observed minimum distance r0}

\begin{figure}[htb]
  \begin{center}
    \includegraphics[width=0.45\textwidth]{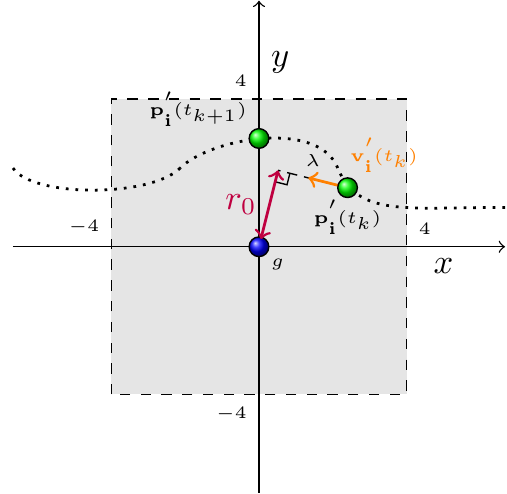}
  \end{center}
  \caption{Illustration of
    computation of the observed minimum distance $r_0$. }
  \label{fig:distance_straight_interpolated}
\end{figure}

The trajectories provided at the ATR pedestrian group data set~\cite{atr2015pedestrian} are derived by the algorithm of Br\v{s}\v{c}i\'c et al.~\cite{brscic2013person}, whose output rate depends on the rate of sensor readings, which may be non-uniform.  In order to make the rate of trajectory samples uniform and also to eliminate the effect of gait and sensor noise, we  re-sampled the trajectories at 2~Hz. However, the new time resolution can be too sparse for the purpose of computing the observed minimum distance. As a remedy to this issue, we propose interpolating the position of an individual  $i$ between two consecutive time steps $t_k$ and $t_{k+1}$ with its own velocity vector at time $t_k$, $\mathbf{v}^{\prime}_i(t_k)$ (see Figure~\ref{fig:distance_straight_interpolated}).

For each time step $t_k$, the smallest distance between $g$ and $i$ can be computed by measuring the distance from the origin to the line passing through $p^{\prime}_i(t_k)$, the position of $i$ at $t_k$ and directed by $\mathbf{v}^{\prime}_i(t_k)$, the velocity of $i$ at $t_k$. We can consider this distance as the actual minimal distance to the group $g$ in the interval $\left[t_k,t_{k+1}\right]$, only if $i$ can reach the position where this distance is observed within $ t_{k+1} - t_k $ seconds (0.5~s in this study). We  denote the signed distance from $p^{\prime}_i(t_k)$ to that position with $\lambda$, which can be computed by taking the projection of the vector from $p^{\prime}_i(t_k)$ to the origin (i.e.\ $-\mathbf{v}^{\prime}_i(t_k)$) onto the unit vector $\frac{\mathbf{v}^{\prime}_i(t_k)}{||\mathbf{v}^{\prime}_i(t_k)||}$,

In explicit terms,
\begin{equation}
  \lambda = \frac{-\mathbf{v}^{\prime}_i(t_k)\cdot\mathbf{p}^{\prime}_i(t_k)}{||\mathbf{v}^{\prime}_i(t_k)||},
\end{equation}

The distance $\lambda$ is traveled by $i$ in $t_{min}$,
\[
  t_{min} = \frac{\lambda}{||\mathbf{v}^{\prime}_i(t_k)||}.
\]
Note that if $t_{min} < t_{k+1} - t_k $, a smaller distance is achieved within the time interval $\left[t_k,t_{k+1}\right]$ than at its initial and final instants ($t_k$ and $t_{k+1}$). Thus, this value is registered as the minimum distance concerning this time interval. This implies that the minimum distance concerning that time interval is achieved at an intermediate instant. Otherwise the lower one of $\left\| \mathbf{p}^{\prime}_i (t_k) \right \|$ and $\left\| \mathbf{p}^{\prime}_i (t_{k+1}) \right \|$ is registered. In this case, if $||\mathbf{p}^{\prime}_i(t_{k})|| < ||\mathbf{p}^{\prime}_i(t_{k+1})||$, it means that $g$ and $i$ are getting further away. If $||\mathbf{p}^{\prime}_i(t_{k})|| > ||\mathbf{p}^{\prime}_i(t_{k+1})||$, it means that $g$ and $i$ are getting closer and that it is highly likely that in the next time interval, they will be even closer. The same operation is carried out for all time intervals and the minimum of all the registered values is used as $r_0$.


\section{Scaling by interpersonal distance}
\label{sec:Scaling by interpersonal distance}


Y\"ucel et al.~\cite{yucel2019identification} showed that interpersonal distance between members of a dyad strongly depends on their social relation. For instance, couples were shown to walk with an interpersonal distance significantly smaller than for other social relations (values of interpersonal distance for various social relations and intensities of interaction are show in Table~\ref{tab:interpersonal_distance}).

This variability may affect the behavior of the individual approaching the group. Namely, an individual $i$ may choose not to intrude a group $g$ due to insufficient space (between its members) or may prefer to intrude  $g$ due to ample space. So the presence or lack of intrusion may simply be due to geometric circumstances and not stemming from social factors (e.g.\ strength of social bonding). This effect is actually better studied by measuring distances using a common unit (i.e.\ meters).

On the other hand, when $i$ does not intrude on $g$, the effect of the social bonding may be better expressed, if distances are measured with respect to the physical size of the group (interpersonal distance between the members), since avoiding a small group at a distance of, e.g.\ 2 meters, may involve a stronger avoidance behavior, if the group  size is smaller.

\begin{table}[!htb]
  \caption{Average interpersonal distance of {groups} annotated with each (a) social relation (in ATC data set) and (b) intensity of interaction (in DIAMOR data set). }
  \label{tab:interpersonal_distance}
  \begin{center}
    \begin{tabular}{cc}
      (a) & (b)
      \\
      \begin{tabular}{lr}
        \toprule
        Social relation & Average interpersonal distance (mm) \\ \midrule
        Couples         & $780$                               \\
        Colleagues      & $891$                               \\
        Family          & $935$                               \\
        Friends         & $835$                               \\ \bottomrule
      \end{tabular}
          &
      \begin{tabular}{lr}
        \toprule
        Intensity of interaction & Average interpersonal distance (mm) \\ \midrule
        0 (no interaction)       & $1115$                              \\
        1                        & $951$                               \\
        2                        & $842$                               \\
        3 (strong interaction)   & $805$                               \\  \bottomrule
      \end{tabular}
    \end{tabular}

  \end{center}
\end{table}


\section{Probability of intrusion}
\label{sec:Probability of intrusion}

\begin{figure}[htb]
  \centering
  \begin{subfigure}[b]{0.48\textwidth}
    \centering
    \includegraphics[width=\textwidth]{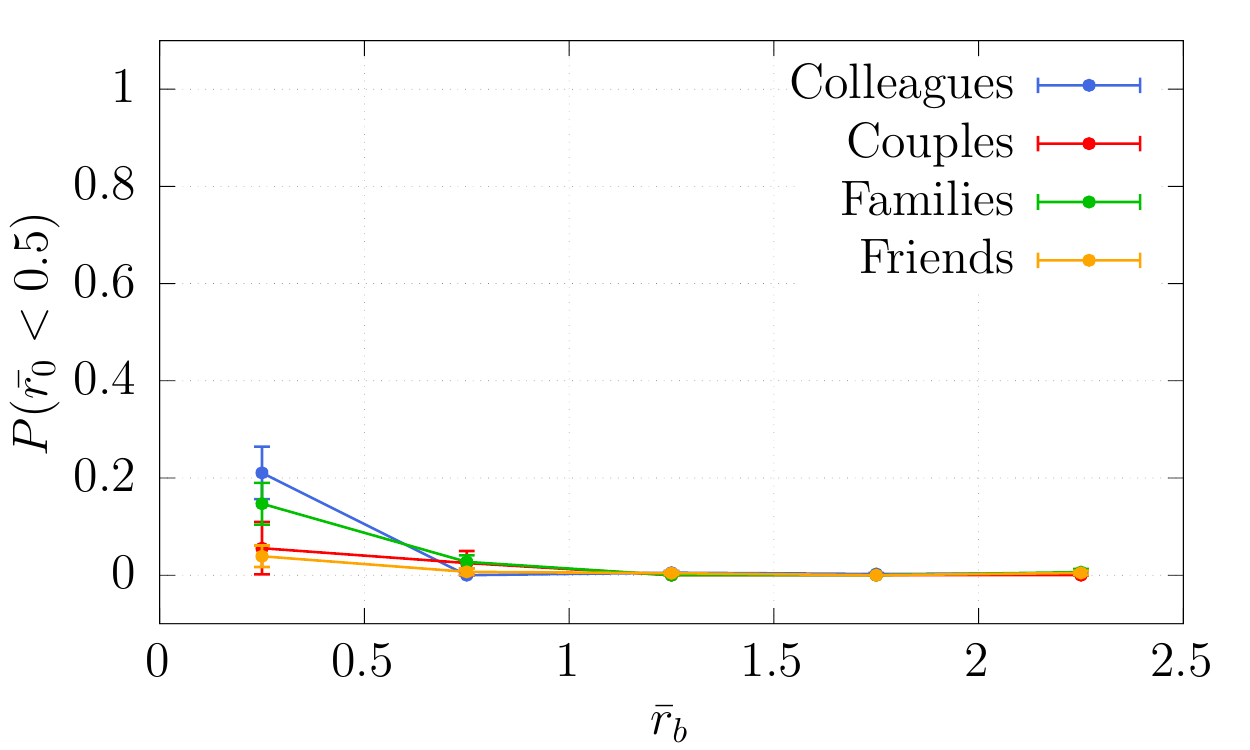}
    \caption{}
    \label{fig:atc_probabilities_05}
  \end{subfigure}
  \hfill
  \begin{subfigure}[b]{0.48\textwidth}
    \centering
    \includegraphics[width=\textwidth]{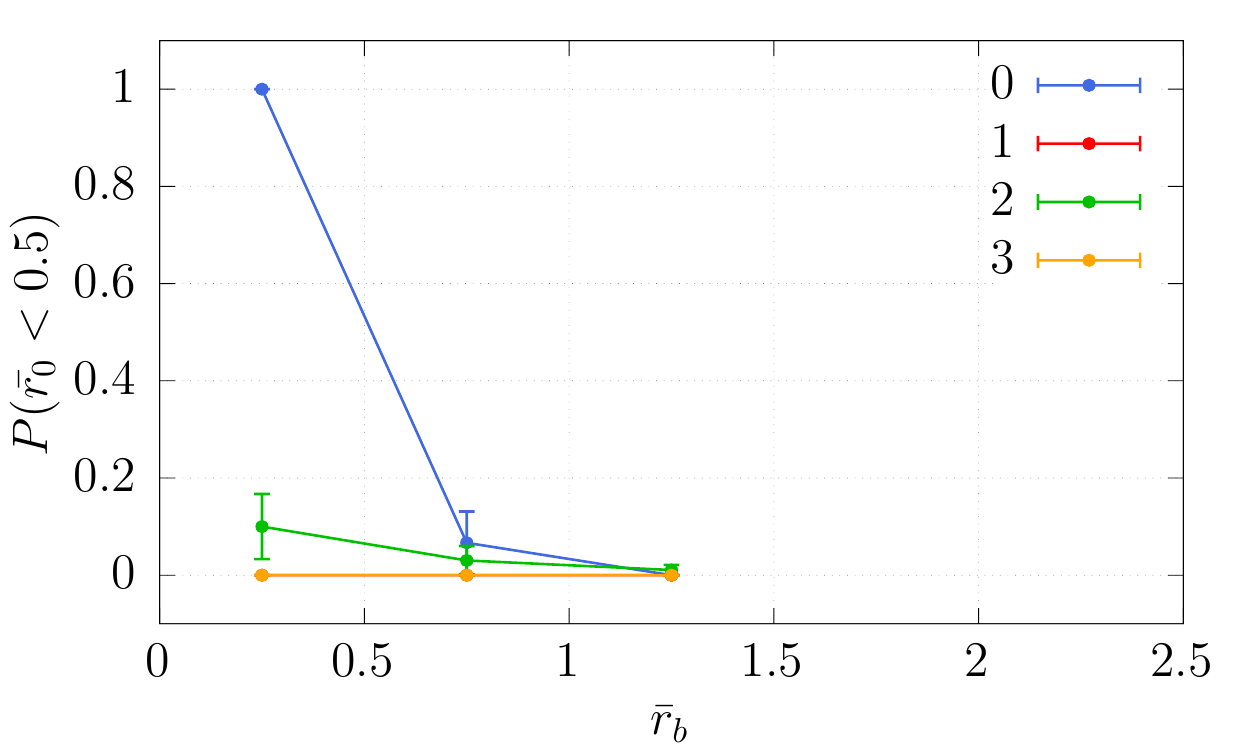}
    \caption{}
    \label{fig:diamor_probabilities_05}
  \end{subfigure}
  \caption{Probability that the distance $\bar{r}_0$ is smaller than $0.5$ for (a) for various social relations and (b) intensities of interaction of the group.}
  \label{fig:probabilities_05}
\end{figure}

\begin{figure}[htb]
  \centering
  \begin{subfigure}[b]{0.48\textwidth}
    \centering
    \includegraphics[width=\textwidth]{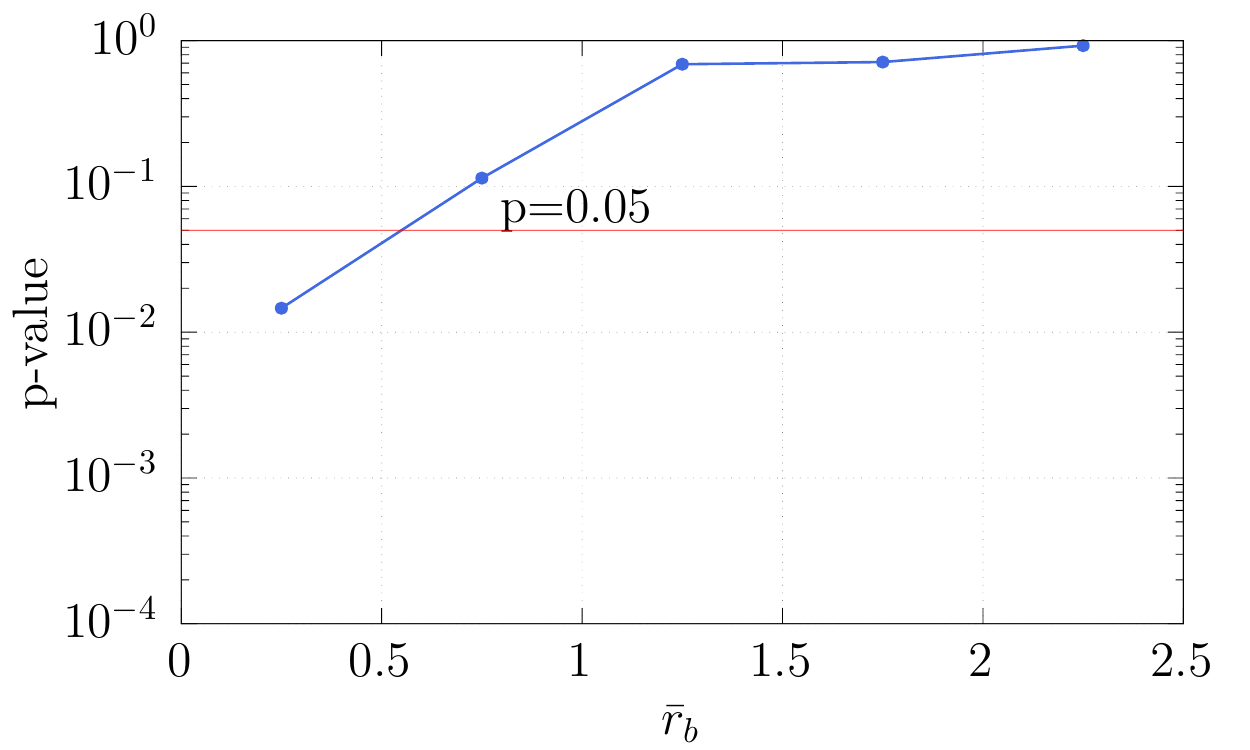}
    \caption{}
    \label{fig:atc_probabilities_pvalues_05}
  \end{subfigure}
  \hfill
  \begin{subfigure}[b]{0.48\textwidth}
    \centering
    \includegraphics[width=\textwidth]{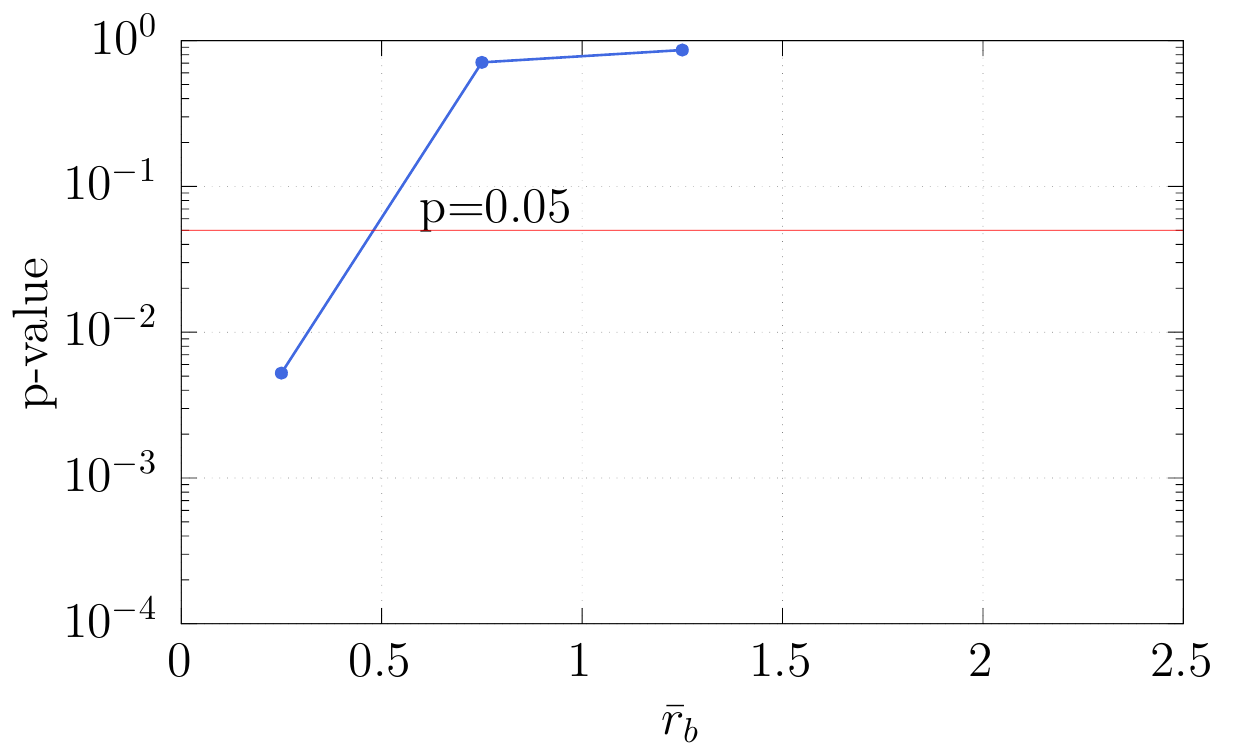}
    \caption{}
    \label{fig:diamor_probabilities_pvalues_05}
  \end{subfigure}
  \caption{Pearson's $\chi^2$ $p$-values for the hypothesis of independence of the frequencies of samples verifying $\bar{r}_0 < 0.5$ and $\bar{r}_0 \geq 0.5$ for (a) for various social relations and (b) intensities of interaction of the group.}
  \label{fig:probabilities_pvalues_05}
\end{figure}

In the main track, to study intrusion, we examined the probability of $\bar{r}_0 < 1$, i.e. that the individual reaches a distance from the group center smaller than the average group interpersonal distance. One could argue that an alternative, if not better, definition of intrusion, can be based on studying the probability of having $\bar{r}_0 < 0.5$, i.e.\ a distance smaller than {\it the average distance of a group member from the group's center}.
For this reason, in this section we perform a similar analysis to the one presented in the main track, but using the value of $0.5$. Namely, Figure~\ref{fig:probabilities_05} shows the probability $P(\bar{r}_0 < 0.5)$ and Figure~\ref{fig:probabilities_pvalues_05} shows the corresponding $p$-values.

Nicely, the probabilities show a similar trend to those given in the main track and confirm our inference that loosely-bonded groups are more likely to be intruded on than strongly-bonded ones.


\section{Linear correction to $r_b$}
\label{sec:Linear correction to rb}

As mentioned in Section~\ref{sec:Window of observation}, we tried to calibrate the window of observation in such a way that the trajectory of the individual should be close to a straight line  for large values of ${r}_b$. If the straight-line distance ${r}_b$  is large, it means that the individual should have enough space to pass comfortably without deviating and we would expect the minimum distance ${r}_0$ to be somewhat similar to ${r}_b$.

Nevertheless, it seems clear that the curved and narrow nature of the ATC environment puts a limit on the applicability of the straight line hypothesis. In particular, as the width of the ATC corridor is comparable to the size of the chosen  window of observation (see Figure~\ref{fig:occupancy_grid}),
we can expect that the environment will pose some constraint on the motion of the pedestrians in particular for large values of $r_b$.

\begin{figure}[htb]
  \begin{center}
    \begin{tabular}{cc}
      \includegraphics[width=0.45\textwidth]{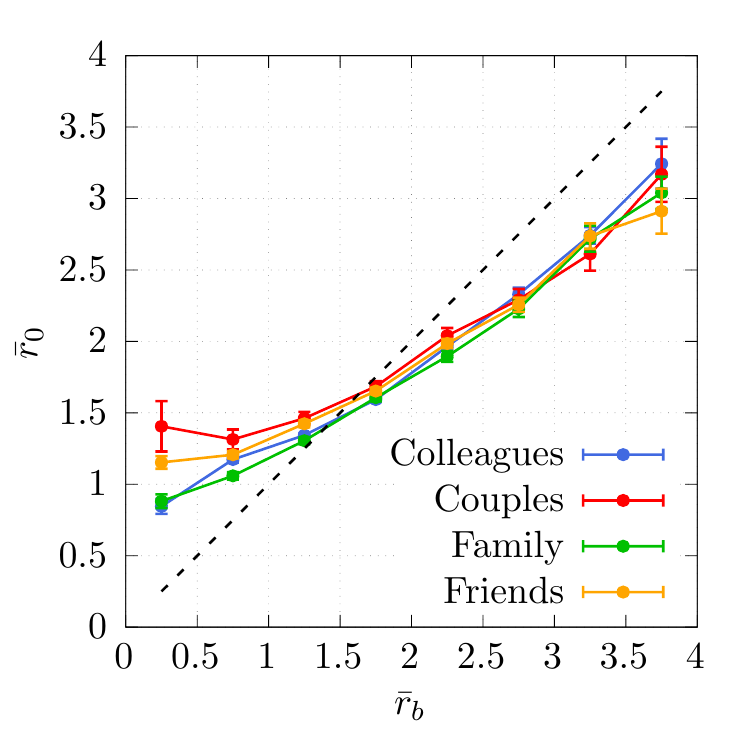}
       &
      \includegraphics[width=0.45\textwidth]{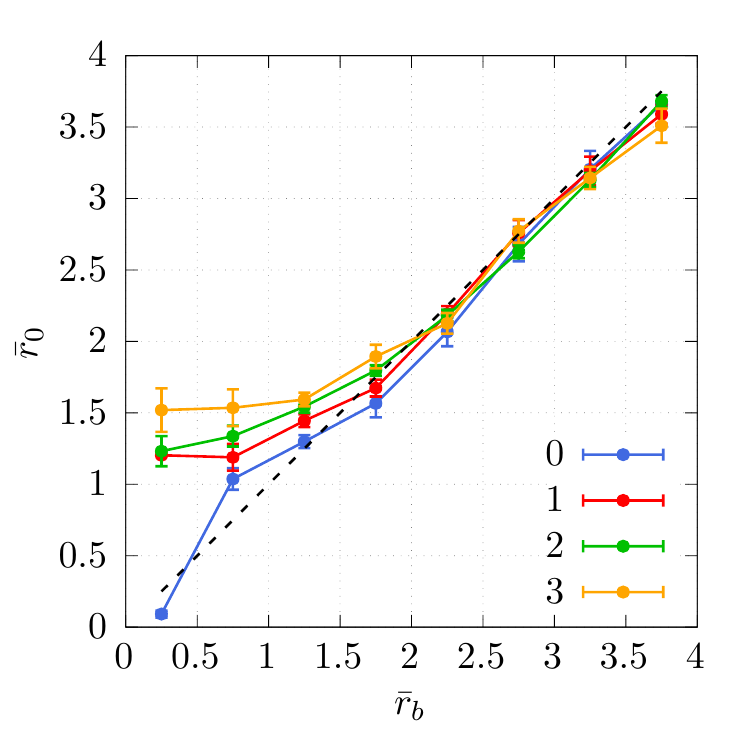}
      \\
      (a)
       &
      (b)
    \end{tabular}
  \end{center}
  \caption{Observed minimum distance $\bar{r}_0$ as a function of the undisturbed straight-line distance $\bar{r}_b$ (a) for various social relations and (b) intensities of interaction of the group. The dashed line corresponds to the $\bar{r}_0=\bar{r}_b$ linear dependence.}
  \label{fig:rb_r0}
\end{figure}

This is indeed confirmed by the data. Namely, by looking at Figure~\ref{fig:rb_r0}-(a) relating to ATC dataset, we observe that the curves are all noticeably offset from the $x=y$ (dashed) line for high values of $\bar{r}_b$. As  $\bar{r}_0$ is  smaller than  $\bar{r}_b$, it seems  as if the individual  \textit{steers towards} the group.

It is not trivial to propose a geometric model for such deviation, due to the relatively complex nature of the ATC environment, but for simplicity's sake we may thus assume the correction on $r_b$ to be linear. Following this hypothesis, we evaluate the impact of  environment geometry by computing an average value of the observed distance $r_0$ (resp. $\bar{r}_0$) for large values of $r_b$ (resp. $\bar{r}_b$) (corresponding to the highest bin in Figure~\ref{fig:rb_r0}), for all groups and individuals. We then compute a correction coefficient $c$ defined as the ratio between the observed average value and the expected value $r_0=r_b$ (as stated above, when $r_b$ is large, we expect no deviation during encounters). Such coefficients are found to be $0.82$ for scaled values and $0.77$ for unscaled values. These coefficients can then be used to multiply the values and alleviate the effect of the curvature of the environment. Specifically, when computing the potential $U'$, we replace the values of $r_b$ (resp. $\bar{r}_b$) by the corrected values $r'_b=c r_b$ (resp. $\bar{r}'_b=c \bar{r}_b$).

\begin{figure}[htb]
  \begin{center}
    \includegraphics[width=0.45\textwidth]{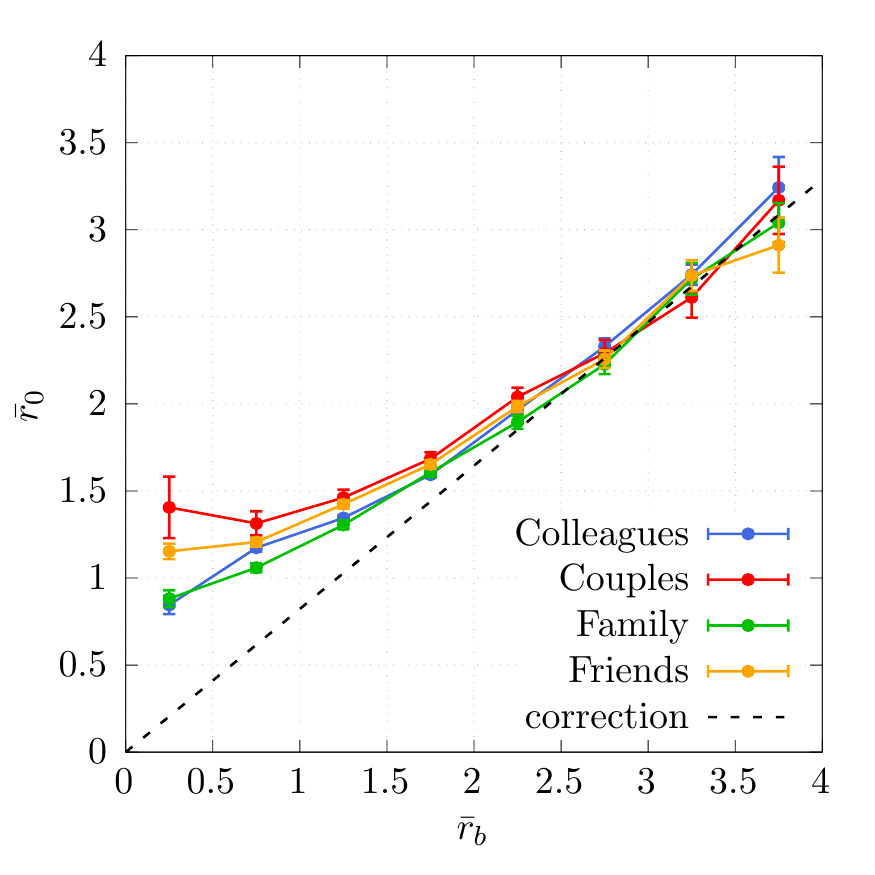}
  \end{center}
  \caption{Observed minimum distance against straight-line distance for various social relations (scaled). The dashed black line corresponds to the linear correction applied. }
  \label{fig:corrections_scaled}
\end{figure}

\begin{figure}[htb]
  \begin{center}
    \begin{tabular}{cc}
      \includegraphics[width=0.45\textwidth]{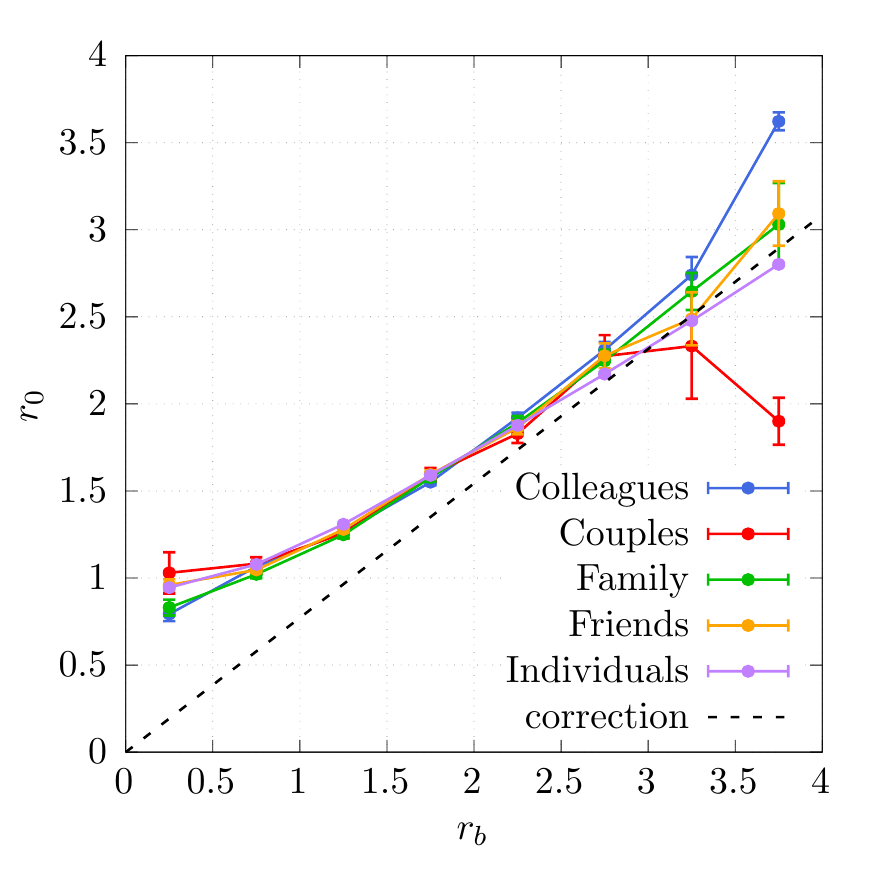}
       &
      \includegraphics[width=0.45\textwidth]{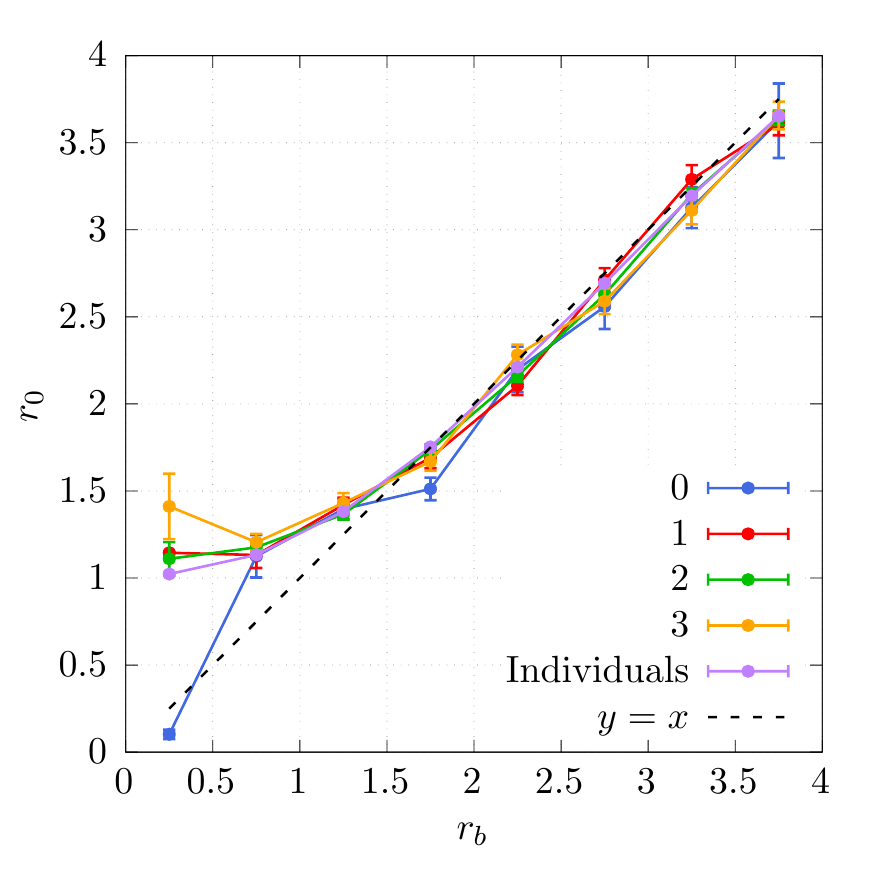}
    \end{tabular}
  \end{center}
  \caption{Observed minimum distance $r_0$ against straight-line distance $r_b$ for (a) various social relations (unscaled) and  (b) groups and intensities of interaction (unscaled). The dashed black line corresponds to (a) the linear correction applied for the ATC dataset and (b) to the line $y=x$ for the DIAMOR dataset.}
  \label{fig:corrections_unscaled}
\end{figure}

In Figure~\ref{fig:corrections_scaled}, we show the scaled  distances for groups with various social relations, along with the line corresponding to the correction coefficient. By definition, the correction  fits the various curves for larger values of $r_b$. For further reference, Figure~\ref{fig:corrections_unscaled} shows the correction coefficient for the unscaled values concerning all groups and individual pedestrians used for the potential of Figure~7 in the main track. The qualitative agreement between the scaled ATC plots and the unscaled DIAMOR ones suggests that
the linear correction used to obtain $r'_b$ and $\bar{r}'_b$ is reasonable.

\section{Justification of using  ANOVA in assessment of statistical significance}
\label{sec:Assumptions for ANOVA stastical tests}

To validate the statistical significance of the differences observed in the relation between $\bar{r}_0$ and $\bar{r}_b$ with regard to the social bonding of the group, we performed analysis of variance tests. The standard one-way ANOVA  requires the data to confirm three conditions, namely (i) independence of  observations, (ii) normality of  residuals and (iii) equality of  variances. In what follows, we verify and discuss these assumptions.

For (i)  independence of  observations, we argue that the independence is naturally verified, since the distances are computed for different pairs of groups and individuals, with each group being classified with only one social bonding characteristic.

Regarding (ii)  normality of  residuals, we carried out D'Agostino's $K^2$ test on the residuals in each bin (i.e.\ the distance from which the average values for that bin was subtracted). We found that in 31\% of the comparisons (i.e.\ 5 out of the 16 bins where the test is performed), the assumption of normality was verified ($p$-value $> 0.05$).

As for  (iii) equality of  variances, we used Levene's test to verify that the various samples have equal variance in each bin. We found that the requirement is satisfied in 75\% of the cases ($p$-value for Levene's test $> 0.05$).

\begin{figure}[htb]
  \begin{center}
    \begin{tabular}{cc}
      \includegraphics[width=0.45\textwidth]{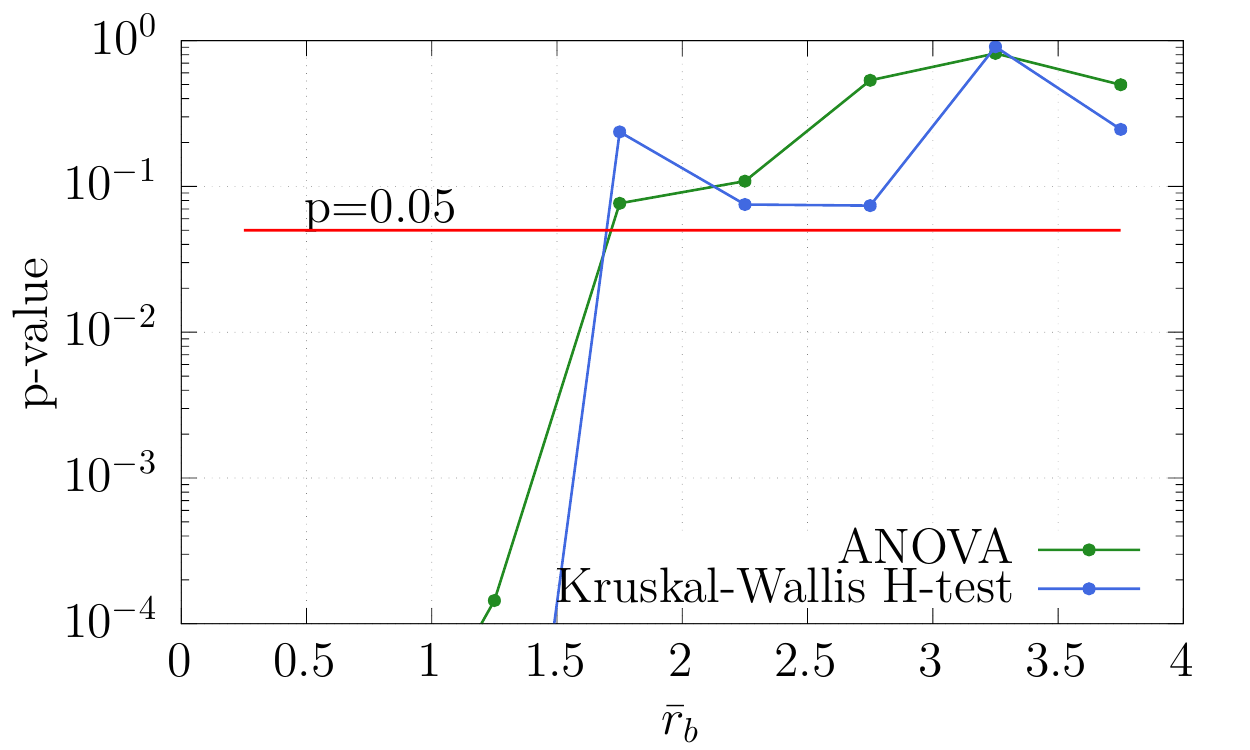}
       &
      \includegraphics[width=0.45\textwidth]{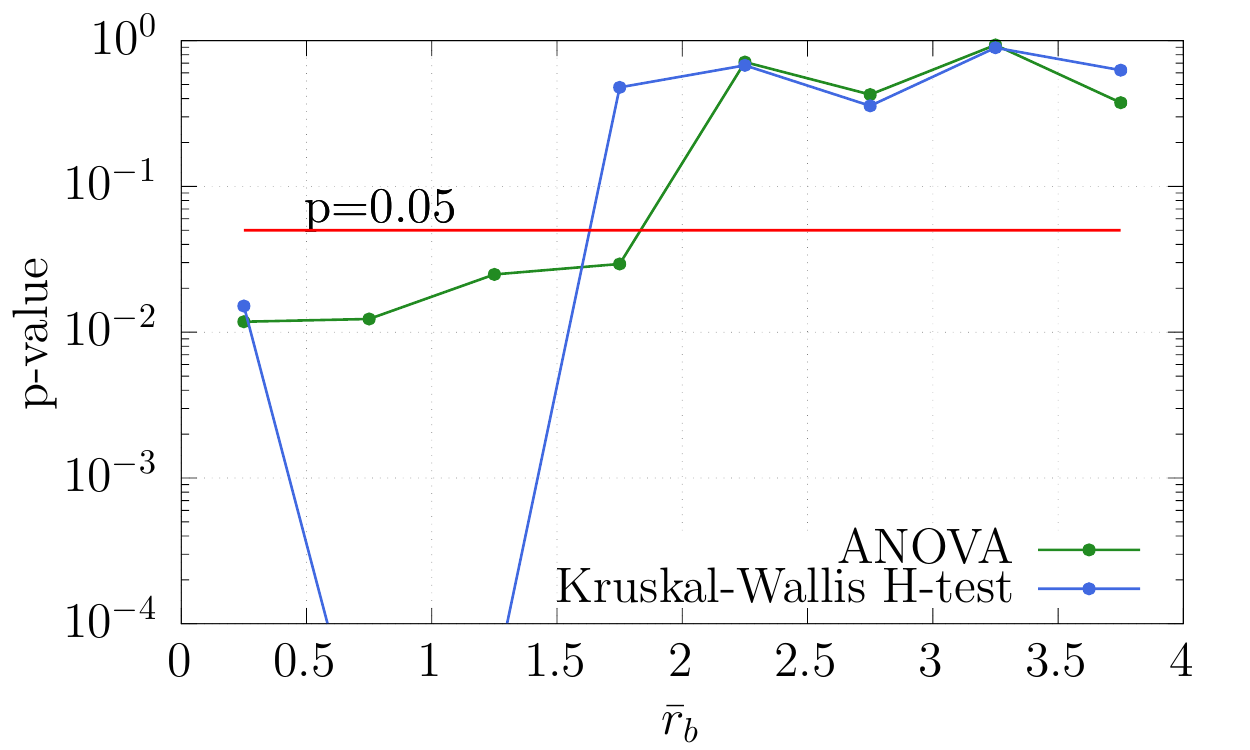}
    \end{tabular}
  \end{center}
  \caption{$p$-values for the ANOVA and Kruskal-Wallis H test of $\bar{r}_0$ (a) for various social relations and (b) intensities of interaction of the dyad. Very low values are not displayed.}
  \label{fig:kruskal_wallis_pvalues}
\end{figure}

From the above, we inferred that normality of the residuals is the requirement which is   not upheld most often,  shedding a doubt on the validity of our conclusions. In that respect, we  employed an additional statistical test, namely Kruskal-Wallis H-test, which is a non-parametric alternative for the one-way ANOVA.  The reason for choosing this test is that it does not require normality of residuals   unlike the standard one-way ANOVA.
The comparison of the results of Kruskal-Wallis H-test to those of ANOVA is presented  in Figure~\ref{fig:kruskal_wallis_pvalues}. From this figure,  one can easily notice that the judgment of statistical significance is not affected by the choice of the test (significance is verified for identical bins). Note that the only bin for which there is a strong difference is $\bar{r}_b \approx 2$ (see Figure~\ref{fig:kruskal_wallis_pvalues}-(b)), which is nevertheless (in both Figure~\ref{fig:kruskal_wallis_pvalues}-(a) and (b)) the transition between significant and insignificant.

\section{Goodness of fit of the model}
\label{sec:Goodness of fit of the model}
Concerning both the scaled and not-scaled distances, the goodness of fit of the proposed models was assessed based on the Akaike information criterion (AIC) and the Kolmogorov–Smirnov (KS) goodness of fit test (see Table~\ref{tab:aic}).

It is hard to  interpret the outcomes of AIC concerning a single model, since it is not bounded (it will tend to $-\infty$ when the residual sum of squares gets smaller). Nonetheless, AIC can be used to compare several models, and thus we use it as such. The fit of the model is seen to be relatively poor for non-interacting  dyads (see Table~\ref{tab:aic}-(b)) compared to other intensities of interaction and social relations. We also note that the fit is generally better, for scaled distances as compared to the unscaled distances (5 models out of 8).

Concerning the Kolmogorov–Smirnov goodness of fit test, the null hypothesis   is that the underlying distributions of the two samples are identical. Our results show that we cannot reject that hypothesis in most cases (note the  high $p$ values in Table~\ref{tab:aic}).

For the intensity of interaction of 0 with a $p$ value of 0.02, it is not surprising that the fit is not good, since there is virtually no effect of groups social interaction at that level, as it can be seen in Figure~5-(b) of the main track.

\begin{table}[!htb]
  \caption{AIC and Kolmogorov-Smirnov $p$-values for the goodness of fit of models describing the collision avoidance potential as a function of the distance $r$ of the form ${k}/{r^\beta}$ for (a) for various social relations and (b) intensities of interaction of the group.}
  \label{tab:aic}
  \begin{center}
    \begin{tabular}{c}
      (a)
      \\
      \begin{tabular}{lrrrr}
        \toprule
        Social relation & \multicolumn{2}{c}{AIC} & \multicolumn{2}{c}{KS $p$-value}                       \\ \midrule
                        & Not-Scaled              & Scaled                           & Not-Scaled & Scaled \\\midrule
        Couples         & -8.51                   & -29.10                           & 0.66       & 1.00   \\
        Colleagues      & -27.10                  & -33.98                           & 0.28       & 0.98   \\
        Family          & -50.76                  & -40.92                           & 0.98       & 0.98   \\
        Friends         & -42.72                  & -39.34                           & 0.98       & 0.98   \\
        Individuals     & -44.73                  & ---                              & 0.98       & ---    \\ \bottomrule
      \end{tabular}
      \\
      (b)
      \\
      \begin{tabular}{lrrrr}
        \toprule
        Intensity of interaction & \multicolumn{2}{c}{AIC} & \multicolumn{2}{c}{KS $p$-value}                       \\ \midrule
                                 & Not-Scaled              & Scaled                           & Not-Scaled & Scaled \\\midrule
        0                        & 54.37                   & 57.90                            & 0.02       & 0.02   \\
        1                        & -27.78                  & -28.82                           & 0.28       & 0.28   \\
        2                        & -42.55                  & -37.55                           & 0.09       & 0.28   \\
        3                        & -18.22                  & -29.45                           & 0.28       & 0.66   \\
        Individuals              & -48.41                  & ---                              & 0.28       & ---    \\ \bottomrule
      \end{tabular}
    \end{tabular}
  \end{center}
\end{table}

\end{document}